\documentclass[pra, aps, twocolumn,amssymb,amsmath,amsbsy, groupedaddress, superscriptaddress]{revtex4-1}
\usepackage{slashed}
\usepackage{graphicx}
\usepackage{float}
\usepackage{subcaption}
\usepackage{enumerate}
\usepackage{chngcntr}
\usepackage[usenames, dvipsnames]{color}
\usepackage{graphics,amsfonts,amsmath, amssymb}
\usepackage{hyperref}
\usepackage{bm}
\usepackage{amsfonts}
\usepackage{dsfont}
\usepackage{lipsum}
\usepackage[utf8]{inputenc}
\usepackage[T1]{fontenc}
\usepackage{bbold}
\usepackage{times}
\usepackage{color}
\hypersetup{backref,
colorlinks=true,
urlcolor=blue,
linkcolor=blue,
linktoc=page,
citecolor=blue}

\everymath{\displaystyle} 

\begin{document}

\title{Multiparameter Quantum Estimation Theory in Quantum Gaussian states}

\author{Lahcen Bakmou}\email{baqmou@gmail.com} \email{lahcen_bakmou@um5.ac.ma}
\affiliation{LPHE-Modeling and Simulation, Faculty of Sciences, Mohammed V University, Rabat, Morocco.}
\author{Mohammed Daoud} \email{m_daoud@hotmail.com}
\affiliation{Department of Physics , Faculty of Sciences, University Ibn Tofail, Kenitra, Morocco.}
\affiliation{LPHE-Modeling and Simulation, Faculty of Sciences, Mohammed V University, Rabat, Morocco.}
\author{Rachid ahl laamara} \email{ahllaamara@gmail.com}
\affiliation{Centre of Physics and Mathematics (CPM), Mohammed V University of Rabat, Rabat, Morocco.}
\affiliation{LPHE-Modeling and Simulation, Faculty of Sciences, Mohammed V University, Rabat, Morocco.}


\begin{abstract}
	
Multiparameter quantum estimation theory aims to determine simultaneously the ultimate precision of all parameters contained in the state of a given quantum system. Determining this ultimate precision depends on the quantum Fisher information matrix (QFIM) which is essential to obtaining the quantum Cramér-Rao bound. This is the main motivation of this work which concerns the computation of the analytical expression of the QFIM. Inspired by the results reported in J. Phys. A 52, 035304 (2019), the general formalism of the multiparameter quantum estimation theory of quantum Gaussian states in terms of their first and second moments is given. We give the analytical formulas of right logarithmic derivative (RLD) and symmetric logarithmic derivative (SLD) operators. Then we derive the general expressions of the corresponding quantum Fisher information matrices. We also derive an explicit expression of the condition which ensures the saturation of the quantum Cramér-Rao bound in estimating several parameters. Finally, we examine some examples to clarify the use of our results.
\end{abstract}

\maketitle

\section{ Introduction}
Besides its fundamental aspects, quantum physics provided us the tools to understand the microscopic world and this understanding has lead to the technological revolution that gives us several solid-state devices.  In the last two decades, it has been theoretically and experimentally shown that quantum mechanics provides the key tools for a modern technological revolution. This type of technology is usually called the quantum technologies \cite{Caves1981, paris2009, toth2014, dowling2015}, among its aspects we mention, quantum communication \cite{briegel1998, duan2001, Gisin2007}, quantum cryptography \cite{ralph1999, bechmann2000,jennewein2000, grosshans2002}, quantum computation \cite{knill2001, lloyd2000, taddei2013} and quantum metrology \cite{giovannetti2006, demkowicz2012, escher2011, pang2014}. The latter constitutes a promising quantum protocol, to enhance the precision of measurements, taking into account the need to discover more sensitive and accurate detectors\cite{giovannetti2004, giovannetti2011, zwierz2010}.

Quantum metrology or quantum estimation theory, was initially proposed by Helstrom \cite{helstrom1976} and Holevo \cite{gudder1985}. Its main goal is to perform high-precision measurements of the parameters specifying a given quantum system. In this sense, quantum metrology aims to develop quantum strategies that allow us to understand the optimal limits of quantum measurements in estimation protocols. The standard limits that fix the ultimate accuracy are known by the quantum Cramér-Rao bounds (QCRB) \cite{gill2005, paris2009}, which always reaches saturation in the case where a single parameter is estimated. On the contrary, it is difficult to saturate this bound in the case of simultaneous estimation of several parameters, due to the incompatibility between the optimal measurements of various estimated parameters \cite{ragy2016, matsumoto2002, vaneph2013, vidrighin2014, crowley2014}. For this, the multiparameter quantum metrology has attracted great interest to generalize certain conditions to saturate the QCRB, and therefore to achieve maximum precision.

In general, to determine the QCRB, it is necessary to compute the quantum Fisher information matrix (QFIM) \cite{paris2009,vsafranek2018simple}. This matrix represents a key ingredient in multiparameter quantum metrology as long as its inverse provides the limits of the maximum precision in the multiparameter estimation. Therefore, the ways to increase the QFIM become an intriguing point in multiparameter protocols enhancement. The QFIM is important for a variety of purposes such as; improvement of the standard frequency \cite{boss2017, albarelli2017, albarelli2018, frowis2014, kessler2014}, estimation of the Unruh-Hawking effect \cite{aspachs2010, huang2018,  liu2019}, magnetic field detection \cite{nair2016,bakmou2019, zhang2014}, applications in thermometry \cite{monras2011, correa2015} and optical interferometry used in the detection of gravitational waves as LIGO \cite{abbott2009} and VIRGO \cite{acernese2014}. In addition,  the QFIM has been also connected to other aspects of quantum mechanics namely, the description of criticality and quantum phase transitions \cite{zanardi2007, zanardi2008, venuti2007}, the quantification of quantum coherence and quantum entanglement \cite{seveso2019, hauke2016, zhang2013, liu2017}. These various potential applications stimulate to develop some theoretical computation techniques to find the QFIM elements. In this context, we present in this paper an analytical method to get the QFIM in bosonic continuous variable systems described by states of Gaussian type.

Recently, Gaussian states that use a continuous variable (CV) systems in the process of quantum information \cite{ferraro2005,braunstein2005, andersen2010} have attracted considerable attention in the literature for two reasons; firstly, for the simplicity of their analytical tools on the theoretical viewpoint owing to, they are described only by the first and second moments, secondly, for its ease to generate and manipulate them experimentally. Indeed, they have several applications in quantum optics \cite{hammerer2010}, optomechanics \cite{tian2010, nunnenkamp2011} and teleportation channel \cite{kim2002, olivares2003, wolf2007}. In addition to that, there is a strong motivation for the Gaussian representation on the remarkable experimental observation in Bose-Einstein condensate \cite{kevrekidis2003, gross2011, wade2016}.

Given the importance of the representation of Gaussian states and the role of multiparameter quantum estimation theory in improving precision measurement, it would be preferable if these two are successfully integrated into a common framework. The goal of our work goes in this direction. We will provide the analytical expression of the central quantities in multiparameter quantum estimation theory, namely the right logarithmic derivative (RLD) and its associated QFIM, as well as the symmetric logarithmic derivative (SLD) and its associated QFIM. Indeed, the most efficient and the most appropriate way to achieve this goal is to use a  phase-space analysis. It must be emphasized that the ideas developed in this work complete some recently obtained results in the literature like for instance. The paper \cite{vsafranek2018} in which the authors derived the quantum Fisher information matrix (QFIM) associated with the symmetric logarithmic derivative (SLD) when the Williamson’s decomposition of the covariance matrix is known. In this paper, we shall provide an easy algorithm to derive the analytical formulas of the quantum Fisher information matrix corresponding to (RLD) and (SLD) simultaneously.  We believe that the results presented here can be adapted to quantum estimation issues involving continuous variables based on the phase-space approach which is was initially proposed in Ref. \cite{monras2013}. It is also interesting to mention that the results obtained in our  work can be adapted  to that obtained in Ref. \cite{genoni2013}

This paper is structured as follows. The second section reviews some basic tools of quantum Gaussian states that are needed for our purpose. Next, we present in  Sec. \ref{sec3} the general framework of the multiparameter quantum estimation theory. In Sec. \ref{sec4} we derive the expressions of RLD and SLD and the corresponding QFIM. We give in Sec. \ref{sec5} some illustrative instance to exemplify the use of our obtained results. Finally, we end this paper with concluding remarks. Technical proofs of computation are provided in the appendices

\section{preliminary for quantum Gaussian states}

Our analysis focuses on the  $N$-mode bosonic CV system described by the creation and annihilation operators $\hat{a}_k^\dag $, ${\hat{a}_k}$ $\left( {k = 1,2,...,N} \right)$ which verify the commutation relations $\left[ {{{\hat a}_j},\hat a_k^\dag } \right] = {\delta _{jk}}$. The Hilbert space for the whole system is the tensor product of infinite-dimensional Fock spaces of $\mathcal{H} = \mathop  \otimes \limits_{k = 1}^N {\mathcal{F}_k}$, so that each mode is covered by the base of the eigenstates of number operator $\hat a_k^\dag {{\hat a}_k}$. The CV systems can be also described by the quadrature operators ${\hat q_k}$, ${\hat p_k}$ that satisfy the commutation relations $\left[ {{{\hat q}_j},{{\hat p}_k}} \right] = 2 i \hspace{0.1cm}{\delta _{jk}}$, with $\hbar  = 2$. These quadratic operators write in terms of $\hat{a}_k^\dag $, ${\hat{a}_k}$ as
\begin{equation}
{\hat q_k} = {{{\hat a}_k} + \hat a_k^ + },   \hspace{1.5cm}   {\hat p_k} = i\left( {\hat a_k^ +  - {{\hat a}_k}} \right).
\end{equation}
The commutation relations between the quadrature operators can be written in a form that is useful for analysis in the phase-space. This is given by
\begin{equation}
\left[ {{{\hat r}_j},{{\hat r}_k}} \right] = 2i \hspace{0.1cm}{\Omega _{jk}},
\end{equation}
where $\mathbf{\hat r }= {\left( {{{\hat q}_1},{{\hat p}_1},...,{{\hat q}_n},{{\hat p}_n}} \right)^T}$ is the vector operators and ${\Omega _{jk}}$ are the elements of the matrix $\Omega$ of dimension $2N \times 2N$,
\begin{equation}
\Omega  = \mathop  \oplus \limits_{k = 1}^n \omega,   \hspace{1.5cm}  \omega  = \left[ {\begin{array}{*{20}{c}}
	0&1\\
	{ - 1}&0
	\end{array}} \right].
\end{equation}
We notice that ${\Omega ^T} = {\Omega ^{ - 1}} =  - \Omega $. In quantum mechanics, the density operator ${\hat \rho }$ encodes all the information of the quantum system. For $N$-mode bosonic CV system,  the density operator describing each mode has an equivalent representation in terms of the quasi-probability distribution defined in the phase-space. This representation is characterized by a function called the characteristic function
\begin{equation}
{\chi _{\hat \rho }}\left( {\bf{r}} \right) = Tr\left[ {{{\hat D}_{ - {\bf{r}}}} \hspace{0.1cm}\hat \rho } \right],
\end{equation}
where $\mathbf{r} = {\left( {{q_1},{p_1},...,{q_n},{p_n}} \right)^T}$ is a vector of $2N$ real coordinates in phase-space and ${\hat D_{ - {\bf{r}}}}$ is the  Weyl operator which is given by 
\begin{equation}
{\hat D_{ - {\bf{r}}}} = {e^{ - i{{\bf{r}}^T}\Omega {\bf{\hat r}}}}.
\end{equation}
Setting  $\mathbf{\tilde r} = \Omega \hspace{0.1cm} \mathbf{r}$, the Weyl operator can be written as follows
\begin{equation}
{{\hat D}_{ - {\bf{r}}}} = {e^{i{{{\bf{\tilde r}}}^T}{\bf{\hat r}}}}. \label{6}
\end{equation}
  The state  ${\hat \rho }$  of a $N$-mode CV system is called  Gaussian state if its characteristic function takes the following form
\begin{equation}
{\chi _{\hat \rho }}\left( \mathbf{r} \right) = exp\left[ { - \frac{1}{4}{{\mathbf{\tilde r}}^T}\sigma \hspace{0.1cm}\mathbf{\tilde r} + i\hspace{0.1cm}{{\mathbf{\tilde r}}^T}\mathbf{d}} \right].\label{7}
\end{equation}
 The characteristic function of the Gaussian states is completely described by two important statistical quantities which are the first and second moments. In particular, the first moment called the displacement vector, is expressed by
\begin{equation}
 \mathbf{d}=\left\langle {\mathbf{\hat r}} \right\rangle=Tr\left[ {\hat \rho \hspace{0.1cm}\mathbf{\hat r}} \right], \label{8}
\end{equation}
and the second moment is the covariance matrix $\sigma$. Its elements are given by
\begin{equation}
{\sigma _{jk}} = \frac{1}{2}Tr\left[ {\hat \rho \left\{ {\Delta {{\hat r}_j},\Delta {{\hat r}_k}} \right\}} \right] = \frac{1}{2}\left\langle {\left\{ {\Delta {{\hat r}_j},\Delta {{\hat r}_k}} \right\}} \right\rangle, \label{9}
\end{equation}
where $\Delta {{\hat r}_j} = {{\hat r}_j} - \left\langle {\hat r_j} \right\rangle$ and the symbol $\left\{ {.,.} \right\}$ represents the notation of anticommutator. The covariance matrix $\sigma$ is  a $2N \times 2N$ real symmetric matrix  defined strictly positive and satisfy the uncertainty principle \cite{simon1994}
\begin{equation}
\sigma  + i \hspace{0.1cm}\Omega  \ge 0. \label{10}
\end{equation}

We now consider a unitary transformation $\hat U = \exp \left( { - i\hat H} \right)$ ( $\hat H$ is the  Hamiltonian of system)  that transforms a state ${{\hat \rho }_{in}}$ into ${{\hat \rho }_{out}}$ as follows
\begin{equation}
{{\hat \rho }_{in}} \to {{\hat \rho }_{out}} = U{{\hat \rho }_{in}}{U^\dag }.
\end{equation}
This transformation is called a Gaussian unitary transformation or Gaussian unitary channel when it preserves the Gaussianity of the quantum state, i.e. it converts a Gaussian state into another Gaussian state. In terms of statistical moments $ \mathbf{d}$ in Eq. (\ref{8}) and $\sigma$ in Eq. (\ref{9}), the action of unitary Gaussian transformation is characterized by the following transformations
\begin{equation}
{\mathbf{d}_{in}} \to {\mathbf{d}_{out}} = S{\mathbf{d}_{in}} + {{\mathbf{r}}}, \hspace{0.5cm}{\sigma _{in}} \to {\sigma _{out}} = S\hspace{0.1cm}{\sigma _{in}}{S^\dag},
\end{equation}
where $\mathbf{r} \in \mathbb{R} {^{2 {N}}}$, and  $S$ is $2N \times 2N$ symplectic real matrix.  More details can be found in  the references \cite{weedbrook2012, braunstein2005}.

\section{Quantum multiparameter estimation theory  \label{sec3}}

Generally, a quantum system is described by a semi-definite positive density operator $\hat \rho$, and all information about this system is encoded in parameters specifying this density operator ${\hat \rho \left( {{\theta _\mu }} \right)}$ such that ${{\theta _\mu } = \left\{ {{\theta _1},...,{\theta _M}} \right\}}$ is the set of parameters contained in the quantum system. Therefore, the main goal of quantum estimation theory is to determine the best possible accuracy in estimating a parameter or several parameters in a metrological protocol. This optimal precision is given by the quantum Cramér-Rao bound (QCRB). The two QCRB most used  are based on RLD and SLD quantum Fisher information matrices \cite{fujiwara1994, genoni2013, gao2014}, where the RLD (right logarithmic derivative) and SLD (symmetric logarithmic derivative ) operators are respectively obtained from the following differential equations (we adopt that ${\partial _{{\theta _\mu }}} = {\raise0.7ex\hbox{$\partial $} \!\mathord{\left/
		{\vphantom {\partial  {\partial {\theta _\mu }}}}\right.\kern-\nulldelimiterspace}
	\!\lower0.7ex\hbox{${\partial {\theta _\mu }}$}}$)
\begin{equation}
{\partial _{{\theta _\mu}}}\hat \rho  = \hat \rho \mathcal{\hat L}_{{\theta _\mu}}^R, \label{13}
\end{equation}
\begin{equation}
{\partial _{{\theta _\mu}}}\hat \rho  = \frac{1}{2}\left\{ {\hat \rho ,\hat L_{{\theta _\mu}}^S} \right\}. \label{14}
\end{equation}
 The quantum Fisher information matrices associated with RLD and SLD are defined respectively by 
\begin{equation}
{{\cal F}_{{\theta _\mu }{\theta _\nu }}} = Tr\left[ {\hat \rho \hat {\cal L}_{{\theta _\mu }}^R\hat {\cal L}{{_{{\theta _\nu }}^R}^\dag }} \right], \label{15}
\end{equation}
\begin{equation}
{H_{{\theta _\mu }{\theta _\nu }}} = \frac{1}{2}Tr\left[ {\hat \rho \left\{ {\hat L_{{\theta _\mu }}^S,\hat L_{{\theta _\nu }}^S} \right\}} \right]. \label{16}
\end{equation}
 The associated QCRB are expressed by
\begin{equation}
\mathbf{Cov\left[ \hat{\theta}  \right]} \ge \frac{{{\mathop{\rm Re}\nolimits} \left[ {{{\cal F}^{ - 1}}} \right] + \left| {{\mathop{\rm Im}\nolimits} \left[ {{{\cal F}^{ - 1}}} \right]} \right|}}{\mathcal{N}}, \label{17}
\end{equation}
\begin{equation}
\mathbf{Cov\left[ \hat{\theta} \right]} \ge \frac{{{H^{ - 1}}}}{\mathcal{N}}, \label{18}
\end{equation}
where $\mathbf{Cov\left[ {\hat \theta } \right]}$ is a covariance matrix defined by $\mathbf{Cov\left[ {{\theta _\mu },{\theta _\nu }} \right]} = E\left( {{\theta _\mu }{\theta _\nu }} \right) - E\left( {{\theta _\mu }} \right)E\left( {{\theta _\nu }} \right)$, the symbol $\left|  \bullet  \right|$ denotes the absolute value of the quantity $\bullet$ and $\mathcal{N}$ is the number the measurements performed. 

In particular, the individual estimation strategy is equivalent to ${\mathcal{F}_{{\theta _\mu }{\theta _\nu }}} = {H_{{\theta _\mu }{\theta _\nu }}} = 0$ when $\mu  \ne \nu $. Therefore, the optimal measure of a parameter can be quantified by the variance, which implies that the Eqs. (\ref{17}) and (\ref{18}) reduce to
\begin{equation}
{\mathop{\rm var}} \left[ {{\theta _\mu }} \right] \ge \frac{{{\rm{Re}}\left[ {F_{{\theta _\mu }{\theta _\mu }}^{ - 1}} \right] + \left| {{\rm{Im}}\left[ {F_{{\theta _\mu }{\theta _\mu }}^{ - 1}} \right]} \right|}}{\mathcal{N}}, \label{19}
\end{equation}
\begin{equation}
{\mathop{\rm var}} \left[ {{\theta _\mu }} \right] \ge \frac{{H_{{\theta _\mu }{\theta _\mu }}^{ - 1}}}{\mathcal{N}}.\label{20}
\end{equation}
The Eqs. (\ref{19}) and (\ref{20}) are always saturated. This saturation corresponds to an optimal measurement of the parameter, and the optimal states forming the projection corresponding to the eigenbasis of SLD. If we apply the trace operator to the two inequalities (\ref{17}) and (\ref{18}), we find that they correspond to the sum of the variances of the estimated parameters
\begin{equation}
\sum\limits_\mu ^M {\rm var\left( {{\theta _\mu }} \right)}  \ge B_R = \frac{{Tr\left[ {{\mathop{\rm Re}\nolimits} \left[ {{{\cal F}^{ - 1}}} \right]} \right] + Tr\left[ {\left| {{\mathop{\rm Im}\nolimits} \left[ {{{\cal F}^{ - 1}}} \right]} \right|} \right]}}{\mathcal{N}}, \label{21}
\end{equation}
\begin{equation}
\sum\limits_\mu ^M {\rm var\left( {{\theta _\mu }} \right)}  \ge B_S = \frac{{Tr\left[ {{H^{ - 1}}} \right]}}{\mathcal{N}}. \label{22}
\end{equation}

In general, the problem remains in scenarios of simultaneous estimation of several parameters. In this case, the limits associated with RLD and SLD can not be saturated because of the incompatibilities between the optimal measurements of the different parameters, i.e. the optimization of the measurement on a parameter can disturb the accuracy of one measure on others. This is the consequence of the noncommutativity of quantum mechanics. On the other hand, the optimal measure for RLD does not always correspond to a positive operator values measurement (POVM) . It is therefore natural to look for the conditions that must be verified in a multi-parameters scenario to saturate these inequalities and finally achieve an optimal measurement. In this context, it is interesting to note that several works on quantum multiparameter estimation theory \cite{ragy2016, matsumoto2002, vaneph2013, vidrighin2014, crowley2014} were devoted to SLD and most of these works showed that the QCRB associated with SLD (\ref{18}), (\ref{22}) can be saturated if and only if
\begin{equation}
Tr\left[ {\hat \rho \left[ {\hat L_{{\theta _\mu }}^S,\hat L_{{\theta _\nu }}^S} \right]} \right] = 0 \label{23}.
\end{equation}
It is simple to see that the condition (\ref{23}) can be equivalently written as
\begin{equation}
{\rm{Im}}\left( {Tr\left[ {\hat \rho \hat L_{{\theta _\mu }}^S\hat L_{{\theta _\nu }}^S} \right]} \right) = 0. \label{24}
\end{equation}
 However, it is natural to ask what is the link between bound $B_R$ associated with RLD and bound $B_S$ associated with SLD. And which of these bounds is more informative and important. Answers to these questions were reported in the references \cite{genoni2013, gao2014} by introducing the so-called  the most informative QCRB ($B_{MI}$) defined by 
\begin{equation}
{B_{MI}} = \max\left\{ {{B_R},{B_S}} \right\}. \label{25}
\end{equation}
Consequently, the determination of the most informative QCRB depends completely on the comparison between the QCRB associated with RLD and the QCRB associated with SLD. For this reason, we introduce the ratio between the two QCRBs, it is defined as follows
	\begin{equation}
	\mathcal{R} = \frac{{{B_S}}}{{{B_R}}},
	\end{equation}
	if $\mathcal{R} < 1$, then $B_{MI}$ corresponds to $B_S$.  If $\mathcal{R} > 1$, then $B_{MI}$ corresponds to $B_R$. In the situation where  $\mathcal{R} = 1$, we will see that $B_{MI} = B_R = B_S$.
\\
Finally, the optimal measures in the multi-parameter protocols can be equated as a single inequality that is given as follows
\begin{equation}
\sum\limits_\mu ^M {{\mathop{\rm var}} \left[ {{\theta _\mu }} \right] \ge \frac{{{B_{MI}}}}{\mathcal{N}}}.
\end{equation}

\vspace{-0.6cm}

\section{Evaluation of RLD and SLD quantum Fisher information  matrices in quantum Gaussian states \label{sec4}}
In this section, we derive the explicit formulas of the RLD and SLD operators in quantum Gaussian states. Using their expression, we determine the analytic expressions of the quantum Fisher information matrices associated with RLD and SLD respectively. To simplify our notations, we adopt in what follows the Einstein’s convention of summation over repeated indices.

\subsection{Evaluation of RLD quantum Fisher information matrix}
It is clear that to determine the elements of the RLD quantum Fisher information matrix,  defined by Eq. (\ref{15}), it is necessary to obtain first the expression of the right logarithmic derivative (RLD) $\mathcal{\hat L_{{\theta _\mu }}}^R$ defined by Eq. (\ref{13}). For a $N$-mode Gaussian state, we consider that RLD must be at most quadratic in the canonical operators:
\begin{equation}
\mathcal{\hat L_{{\theta _\mu }}}^R = {{\cal L}^R}^{\left( 0 \right)} + {\cal L}_l^{R\left( 1 \right)}{\hat r_l} + {\cal L}_{jk}^{R\left( 2 \right)}{\hat r_j}{\hat r_k}, \label{27}
\end{equation}
where $\mathbf{\hat r }= {\left( {{{\hat q}_1},{{\hat p}_1},...,{{\hat q}_n},{{\hat p}_n}} \right)^T}$ is the vector of canonical operators,  ${\mathcal{L}^R}^{\left( 0 \right)} \in \mathbb{C}$, ${\mathbf{\mathcal{L}}^R}^{\left( 1 \right)} $ is a vector in $ \mathbb{C}^{2N}$ and ${\mathcal{{ {L}}}^R}^{\left( 2 \right)}$ is $2N \times 2N$  complex matrix.

For a given set of the parameters $\theta _\mu$, we prove in Appendix \ref{app:A} that the quantities $\hat { \mathcal{L}}_{{\theta _\mu }}^{R\left( 0 \right)}$, \hspace{0.1cm} $\hat { \mathcal{L}}_{{\theta _\mu }}^{R\left( 1 \right)}$ and $\hat { \mathcal{L}}_{{\theta _\mu }}^{R\left( 2 \right)}$ in Eq. (\ref{27}) can be written respectively as follows
\begin{equation}
 {\cal L}_{{\theta _\mu }}^{R\left( 0 \right)} =  - \frac{1}{2}Tr\left[ {{\Gamma _ + }\hat {\mathcal{L}}_{{\theta _\mu }}^{R\left( 2 \right)}} \right] - {{\bf{d}}^T}\hat {\mathcal L}_{{\theta _\mu }}^{R\left( 1 \right)} - {{\bf{d}}^T}\hat {\mathcal{L}}_{{\theta _\mu }}^{R\left( 2 \right)}{\bf{d}},
\end{equation}
\begin{equation}
\hat{ \mathcal{L}}_{{\theta _\mu }}^{R\left( 1 \right)} = 2 \hspace{0.1cm}\Gamma _ + ^{ - 1}\hspace{0.1cm}{\partial _{{\theta _\mu }}}{\bf{d}} - 2\hspace{0.1cm}\hat{ \mathcal{L}}_{{\theta _\mu }}^{R\left( 2 \right)}{\bf{d}},
\end{equation}
\begin{equation}
	\mathtt{vec}\left[ \hat { \mathcal{L}}_{{\theta _\mu }}^{R\left( 2 \right)} \right] = {\left( {\Gamma ^\dag \otimes {\Gamma }} \right)^{ +}}\mathtt{vec}\left[ {{\partial _{{\theta _\mu }}}\sigma } \right],
\end{equation}
where ${\Gamma} = \sigma  + i\hspace{0.1cm}\Omega $ and $\mathtt{vec}\left[A\right]$ denotes the vectorization of a matrix $A$  which defined for any $p \times p$ real or complex matrix $A$ as;
$\mathtt{vec}\left[ A \right] = {\left( {{a_{11}},...,{a_{p1}},{a_{12}},...,{a_{p2}},...,{a_{1p}},...,{a_{pp}}} \right)^T}$.
Inserting the expression of the right logarithmic derivative (RLD) into Eq. (\ref{15}) (the calculation details are found in Appendix \ref{App:B}) we find the RLD quantum Fisher information matrix 
\begin{equation}
{{\mathcal{F}}_{{\theta _\mu }{\theta _\nu }}} = \frac{1}{2}\mathtt{vec}{\left[ {{\partial _{{\theta _\mu }}}\sigma } \right]^\dag }{\Sigma ^ + }\mathtt{vec}\left[ {{\partial _{{\theta _\nu }}}\sigma } \right] + 2{\partial _{{\theta _\mu }}}{\mathbf{d}^T}\hspace{0.1cm}\Gamma ^{+}\hspace{0.1cm}{\partial _{{\theta _\nu }}}\mathbf{d}, \label{31}
\end{equation}
where ${\Sigma ^ + } = \left( {{\Gamma ^\dag } \otimes \Gamma } \right)^+$ and the index "+" denotes the Moore-Penrose pseudoinverse which is a generalization of the inverse matrix \cite{penrose1955, ben2003} that be calculated using the Tikhonov regularization \cite{golub1996}: ${A^ + } = \mathop {\lim }\limits_{\delta  \searrow 0} \left( {{A^\dag }{{\left( {A{A^\dag } + \delta I} \right)}^{ - 1}}} \right) = \mathop {\lim }\limits_{\delta  \searrow 0} \left( {{{\left( {{A^\dag }A + \delta I} \right)}^{ - 1}}{A^\dag }} \right)$. These limits exist even if $A^{-1}$ does not exist.

We note that if $\Gamma$ is  invertible (non-singular) the RLD quantum Fisher information matrix can be expressed as
\begin{equation}
{{\mathcal{F}}_{{\theta _\mu }{\theta _\nu }}} = \frac{1}{2}\mathtt{vec}{\left[ {{\partial _{{\theta _\mu }}}\sigma } \right]^\dag }{\Sigma ^{ - 1}}\mathtt{vec}\left[ {{\partial _{{\theta _\nu }}}\sigma } \right] + 2{\partial _{{\theta _\mu }}}{\mathbf{d}^T}\hspace{0.1cm}\Gamma ^{-1}\hspace{0.1cm}{\partial _{{\theta _\nu }}}\mathbf{d}. \label{32}
\end{equation}
In this case, the Moore-Penrose pseudoinverse of \hspace{0.1cm}$\Gamma$ coincides with its inverse.

\subsection{Evaluation of SLD quantum Fisher information  matrix }
Similarly, the expression of the SLD quantum Fisher information matrix requires the explicit formula of the symmetric logarithmic derivative (SLD) $\hat L_{{\theta _\mu }}^S$ defined by Eq. (\ref{14}). For this end, we also write the SLD as a quadratic form in canonical operators:
\begin{equation}
\hat L_{{\theta _\mu }}^S = {L^S}^{\left( 0 \right)} + L_l^{S\left( 1 \right)}{\hat r_l} + L_{jk}^{S\left( 2 \right)}{\hat r_j}{\hat r_k}, \label{33}
\end{equation}\\
with $L_{{\theta _\mu }}^{S\left( 0 \right)} \in \mathbb{R} $, $ \hat L_{{\theta _\mu }}^{S\left( 1 \right)} \in {\mathbb{R}^{2N}}$ and $\hat L_{{\theta _\mu }}^{S\left( 2 \right)}$ is $2N \times 2N$ real symmetric matrix. These quantities are given respectively by the following expressions (more details are given in Appendix \ref{App:C})
\begin{equation}
L_{{\theta _\mu }}^{S\left( 0 \right)} =  - \frac{1}{2}Tr\left[ {\sigma \hat L_{{\theta _\mu }}^{S\left( 2 \right)}} \right] - {{\bf{d}}^T}\hat L_{{\theta _\mu }}^{S\left( 1 \right)} - {{\bf{d}}^T} \hat L_{{\theta _\mu }}^{S\left( 2 \right)}{\bf{d}},\label{35}
\end{equation}
\begin{equation}
\hat L_{{\theta _\mu }}^{S\left( 1 \right)} = 2 \hspace{0.1cm}{\sigma ^{ - 1}}{\partial _{{\theta _\mu }}}{\bf{d}} - 2 \hspace{0.1cm}\hat L_{{\theta _\mu }}^{S\left( 2 \right)}{\bf{d}},\label{361}
\end{equation}
\begin{equation}
\mathtt{vec}\left[ {\hat L_{{\theta _\mu }}^{S\left( 2 \right)}} \right] = {\left( {{\sigma ^\dag} \otimes \sigma  + \Omega  \otimes \Omega } \right)^{+}}\mathtt{vec}\left[ {{\partial _{{\theta _\mu }}}\sigma } \right]. \label{371}
\end{equation}
Thus, inserting the expression of SLD in Eq. (\ref{16}), one gets the elements of the SLD quantum Fisher information matrix (see Appendix \ref{App:D}).
\begin{equation}
{H_{{\theta _\mu }{\theta _\nu }}} = \frac{1}{2}\mathtt{vec}{\left[ {{\partial _{{\theta _\mu }}}\sigma } \right]^\dag }{{\mathcal M}^{+}}\mathtt{vec}\left[ {{\partial _{{\theta _\nu }}}\sigma } \right] + 2{\partial _{{\theta _\mu }}}{{\bf{d}}^T} \hspace{0.1cm} \sigma ^{-1} \hspace{0.1cm}{\partial _{{\theta _\nu }}}{\bf{d}},\label{37}
\end{equation}
where $\mathcal{M} = \left( {{\sigma ^\dag} \otimes \sigma  + \Omega  \otimes \Omega } \right)$. In the case where $\mathcal{M}$ is invertible, the SLD quantum Fisher information matrix can be calculated as  
\begin{equation}
{H_{{\theta _\mu }{\theta _\nu }}} = \frac{1}{2}\mathtt{vec}{\left[ {{\partial _{{\theta _\mu }}}\sigma } \right]^\dag }{{\mathcal M}^{-1}}\mathtt{vec}\left[ {{\partial _{{\theta _\nu }}}\sigma } \right] + 2{\partial _{{\theta _\mu }}}{{\bf{d}}^T} \hspace{0.1cm} \sigma ^{-1} \hspace{0.1cm}{\partial _{{\theta _\nu }}}{\bf{d}}.\label{38}
\end{equation}
According to Ref. \cite{nichols2018}, the saturation condition of the quantum Cramér-Rao bound (\ref{24}) is expressed in the phase-space as 
\begin{align}\label{39}
{\rm{Im}}\left( {Tr\left[ {\hat \rho \hat L_{{\theta _\mu }}^S\hat L_{{\theta _\nu }}^S} \right]} \right) = 2\hspace{0.1cm}Tr\left[ {\sigma \hat L_{{\theta _\mu }}^{S\left( 2 \right)}\Omega \hat L_{{\theta _\nu }}^{S\left( 2 \right)}} \right] +\\ \notag
 2\hspace{0.1cm}{\partial _{{\theta _\mu }}}{\mathbf{d}^T}{\sigma ^{ - 1}}\Omega {\sigma ^{ - 1}}{\partial _{{\theta _\nu }}}\mathbf{d}.
\end{align}
These results can be rewritten in a compact form using the notations introduced here above. We thus, consider the following relations
\begin{equation}
Tr\left[ {{A^\dag }B} \right] = \mathtt{vec}{\left[ A \right]^\dag }\mathtt{vec}\left[ B \right], \label{40}
\end{equation}
\begin{equation}
\mathtt{vec}\left[ {AB} \right] = \left( {I \otimes A} \right)\mathtt{vec}\left[ B \right] = \left( {{B^\dag } \otimes I} \right)\mathtt{vec}\left[ A \right], \label{41}
\end{equation}
\begin{equation}
\left( {A \otimes B} \right)\left( {C \otimes D} \right) = AC \otimes BD, \label{42}
\end{equation}
From Eqs. (\ref{40}), (\ref{41}) and (\ref{42}), one has
\begin{align}
Tr\left[ {{{\left( {AD} \right)}^\dag }BC} \right] &= \mathtt{vec}{\left[ {AD} \right]^\dag }\mathtt{vec}\left[ {BC} \right]\\ \notag&
= \mathtt{vec}{\left[ A \right]^\dag }\left( {D \otimes I} \right)\left( {I \otimes B} \right)\mathtt{vec}\left[ C \right],
\end{align}
\begin{equation}
Tr\left[ {{A^\dag }BC{D^\dag }} \right] = \mathtt{vec}{\left[ A \right]^\dag }\left( {D \otimes B} \right)\mathtt{vec}\left[ C \right].
\end{equation}
Using the last equation, we find that the first term of Eq. (\ref{39}) can be written as
{\small \begin{align*}
{\small Tr\left[ {L_{{\theta _\mu }}^{S\left( 2 \right)}\Omega L_{{\theta _\nu }}^{S\left( 2 \right)}\sigma } \right]} &= \mathtt{vec}{\left[ {L_{{\theta _\mu }}^{S\left( 2 \right)}} \right]^\dag }\left( {\sigma  \otimes \Omega } \right)\mathtt{vec}\left[ {L_{{\theta _\nu }}^{S\left( 2 \right)}} \right]\\ 
\notag&
  =  \mathtt{vec}{\left[ {{\partial _{{\theta _\mu }}}\sigma } \right]^\dag }{{\cal M}^{+}}\left( {\sigma  \otimes \Omega } \right){{\cal M}^{+}}\mathtt{vec}\left[ {{\partial _{{\theta _\nu }}}\sigma } \right].
\end{align*}}
The last equality follows from Eq. (\ref{371}). Finally, the expression of the saturation condition of the quantum Cramér-Rao bound in terms of $\mathbf{d}$, $\sigma$, and their  derivative with respect to the estimated parameters can be derived as
{\small \begin{align}
\notag
{\rm{Im}}\left( {Tr\left[ {\hat \rho \hat L_{{\theta _\mu }}^S\hat L_{{\theta _\nu }}^S} \right]} \right) &= 2 \mathtt{vec}{\left[ {{\partial _{{\theta _\mu }}}\sigma } \right]^\dag }{{\cal M}^{+}}\left( {\sigma  \otimes \Omega } \right){{\cal M}^{+}}\mathtt{vec}\left[ {{\partial _{{\theta _\nu }}}\sigma } \right] \\ & 
+2{\partial _{{\theta _\mu }}}{{\bf{d}}^T} \hspace{0.1cm}{\sigma ^{-1}}\Omega \hspace{0.1cm} {\sigma ^{-1}}{\partial _{{\theta _\nu }}}{\bf{d}}. \label{45}
\end{align}}
When the matrix $\mathcal{M}$ is invertible,  $\mathcal{M}^+$ can be replaced by $\mathcal{M}^{-1}$ in the last equation.\\
 Eqs. (\ref{35}, \ref{361}, \ref{371}, \ref{38}, \ref{45}) are identical to the Eqs. (9, 8, 11) of Ref. \cite{vsafranek2018} they obtained by a different method.
\section{Application \label{sec5}}

In this section, we treat some protocols of multiparameter quantum Gaussian metrology. The first example can be considered as an illustration of the validity and the usefulness  of our results according to the results reported Ref. \cite{genoni2013}. The second example concerns the estimation of the parameters: squeezing parameter $r$ and phase rotation $\varphi$ when we take the thermal state and coherent state as the inputs states evolving under a Gaussian channel (squeezing and rotation channel)

\subsection{Estimation of two parameters of a displacement operator}
 We first consider the estimation of the two parameters $q_0$ and $p_0$ of the displacement operator $\hat D\left( {{q_0},{p_0}} \right) = \exp \left( {i{p_0}\hat q - i{q_0}\hat p} \right)$ with a measurement on the displaced state ${\rho _{out}} = \hat D\left( {{q_0},{p_0}} \right){\rho _{in}}{\hat D^\dag }\left( {{q_0},{p_0}} \right)$. We take the single-mode Gaussian state following ${\rho _{in}} = \hat S\left( r \right){\rho _{th}}\hat S{\left( r \right)^\dag }$ where $\hat S\left( r \right) = \exp \left( {\frac{r}{2}\left( {{{\hat a}^2} - {{\hat a}^{\dag 2}}} \right)} \right)$ denotes the single-mode squeezing operator and $\rho_{th}$ is the thermal state given by
\begin{equation}
{\rho _{th}} = \sum\limits_{n = 0}^{ + \infty } {\frac{{{{\bar n}^n}}}{{{{\left( {\bar n + 1} \right)}^{n + 1}}}}\left| n \right\rangle \left\langle n \right|}, \label{46}
\end{equation}
where $\bar n = \left\langle {{a^\dag }a} \right\rangle $ is the mean number of photon. The first and second moments of the output state are give by
\begin{equation}
{\mathbf{d}_{out}} = \left[ {\begin{array}{*{20}{c}}
	{{q_0}}\\
	{{p_0}}
	\end{array}} \right],  \hspace{1cm} {\sigma _{out}} = \left( {2\bar n + 1} \right)\left[ {\begin{array}{*{20}{c}}
	{{e^{ - 2r}}}&0\\
	0&{{e^{2r}}}
	\end{array}} \right].
\end{equation}
The RLD quantum Fisher information matrix is calculated from Eq. (\ref{32}). It has the form
\begin{equation}
\mathcal{F} = \left[ {\begin{array}{*{20}{c}}
	{\frac{{2\left( {2\bar n + 1} \right){e^{2r}}}}{{{{\left( {2\bar n + 1} \right)}^2} - 1}}}&{\frac{{ - 2i}}{{{{\left( {2\bar n + 1} \right)}^2} - 1}}}\\
	{\frac{{2i}}{{{{\left( {2\bar n + 1} \right)}^2} - 1}}}&{\frac{{2\left( {2\bar n + 1} \right){e^{ - 2r}}}}{{{{\left( {2\bar n + 1} \right)}^2} - 1}}}
	\end{array}} \right].
\end{equation}
Similarly, the SLD quantum Fisher information is calculated from Eq. (\ref{38}). It is given by
\begin{equation}
H = \left[ {\begin{array}{*{20}{c}}
	{\frac{{2\ {e^{2r}}}}{{\left( {2\bar n + 1} \right)}}}&0\\
	0&{\frac{{2{e^{ - 2r}}}}{{\left( {2\bar n + 1} \right)}}}
	\end{array}} \right].
\end{equation}
The two bounds $B_R$ and $B_S$ can be evaluated from Eqs. (\ref{21}), (\ref{22}) as
\begin{equation}
{B_R} = \left( {2\bar n + 1} \right)\cosh \left( {2r} \right) + 1, \hspace{0.6cm} {B_S} = \left( {2\bar n + 1} \right)\cosh \left( {2r} \right).
\end{equation}
Obviously, the most informative quantum Cramér-Rao bound $B_{MI}$ (\ref{25}) in this case is given by $B_R$:
\begin{equation}
{B_{MI}} = \left( {2\bar n + 1} \right)\cosh \left( {2r} \right) + 1.
\end{equation}
This result coincides with the quantum Cramér-Rao bound that obtained in Ref. \cite{genoni2013}. This confirms the validity  of the formalism developed in this paper for the protocols of multiparameter quantum metrology involving Gaussian states.

\subsection{Estimation of two parameters $r$ and $\varphi $ contained in squeezing and rotation operators}
The second illustration concerns the joint estimation of two parameters: squeezing parameter $r$ and phase rotation $\varphi$, when we consider the thermal state (\ref{46}) as the initial probe state (input state). We assume that this state evolves in squeezing and rotation channels, which transforms the input state into
\begin{equation}
{\rho _{out}} =\hat R\left( \varphi  \right) \hat S\left( r \right){\rho _{th}} \hat S{\left( r \right)^\dag } \hat R{\left( \varphi  \right)^\dag }.
\end{equation}
The symplectic transformations corresponding to this channel are given by
\begin{equation}
\hat R\left( \varphi  \right) = \left[ {\begin{array}{*{20}{c}}
	{\cos \varphi }&{\sin \varphi }\\
	{ - \sin \varphi }&{\cos \varphi }
	\end{array}} \right], \hspace{0.5cm}\hat S\left( r \right) = \left[ {\begin{array}{*{20}{c}}
	{{e^{ - r}}}&0\\
	0&{{e^r}}
	\end{array}} \right], \label{53}
\end{equation}
which leads to the following moments for the output state\vspace{-0.1cm}
\begin{equation}
{\mathbf{d}_{out}} = \hat R\left( \varphi  \right)\hat S\left( r \right){\mathbf{d}_{in}}, \hspace{0.3cm}{\sigma _{out}} = \hat R\left( \varphi  \right)\hat S\left( r \right){\sigma _{in}}\hat S{\left( r \right)^\dag }\hat R{\left( \varphi  \right)^\dag },
\end{equation}
where $\mathbf{d}_{in}$ and $\sigma _{in}$ are the first and second moments of $\rho_{th}$, and they are given by \vspace{-0.15cm}
\begin{equation}
{{\bf{d}}_{in}} = \left[ {\begin{array}{*{20}{c}}
	0\\
	0
	\end{array}} \right], \hspace{1.5cm} {\sigma _{in}} = \left( {2\bar n + 1} \right)\mathbb{1}.
\end{equation}
Now, to compute the RLD quantum Fisher information matrix (\ref{15}), one needs the following expressions
		\begin{equation}
	\mathtt{vec}\left[ {{\partial _\varphi }{\sigma _{out}}} \right] =2\left( {2\bar n + 1} \right)\sinh 2r\left[ {\begin{array}{*{20}{c}}
		{\sin 2\varphi }\\
		{\cos 2\varphi }\\
		{\cos 2\varphi }\\
		{ - \sin 2\varphi }
		\end{array}} \right], \label{56}
	\end{equation}\vspace{-0.15cm}
\begin{equation}
	\mathtt{vec}\left[ {{\partial _r}{\sigma _{out}}} \right] = 2\left( {2\bar n + 1} \right)\left[ {\begin{array}{*{20}{c}}
		{{{\sin }^2}\varphi {e^{2r}} - {{\cos }^2}\varphi {e^{ - 2r}}}\\
		{\sin 2\varphi \cosh 2r}\\
		{\sin 2\varphi \cosh 2r}\\
		{{{\cos }^2}\varphi {e^{2r}} - {{\sin }^2}\varphi {e^{ - 2r}}}
		\end{array}} \right]. \label{57}
	\end{equation}
It easy to verify that $\Gamma $ is invertible, so that ${\Gamma ^ + } = {\Gamma ^{ - 1}}$. Thus one gets 
	\begin{widetext}	
		\begin{equation}
{\Gamma ^{ - 1}} =\left[ {\begin{array}{*{20}{c}}
	{\frac{{\left( {1 + 2\bar n} \right)\left( {2\left( {\cosh \left[ {2r} \right] + \cos \left[ {2\varphi } \right]\sinh \left[ {2r} \right]} \right)} \right)}}{{8\bar n\left( {1 + \bar n} \right)}}}&{ - \frac{{{\rm{i}} + \left( {1 + 2\bar n} \right)\sin \left[ {2\varphi } \right]\sinh \left[ {2r} \right]}}{{4\bar n\left( {1 + \bar n} \right)}}}\\
	{\frac{{{\rm{i}} - \left( {1 + 2\bar n} \right)\sin \left[ {2\varphi } \right]\sinh \left[ {2r} \right]}}{{4\bar n\left( {1 + \bar n} \right)}}}&{\frac{{{{\rm{e}}^{ - 2r}}\left( {1 + 2\bar n} \right)\left( {1 + \cos \left[ {2\varphi } \right] + 2{{\rm{e}}^{4r}}\sin {{\left[ \varphi  \right]}^2}} \right)}}{{8\bar n\left( {1 + \bar n} \right)}}}
	\end{array}} \right]. \label{58}
	\end{equation}
\end{widetext}
The RLD quantum Fisher information matrix, can be calculated from Eq. (\ref{32}) as
{\small  \begin{equation}
\mathcal{F} = \frac{1}{2}\left[ {\begin{array}{*{20}{c}}
	{\mathtt{vec}{{\left[ {{\partial _r}\sigma } \right]}^\dag }{\Sigma ^{ - 1}}\mathtt{vec}\left[ {{\partial _r}\sigma } \right]}&{\mathtt{vec}{{\left[ {{\partial _r}\sigma } \right]}^\dag }{\Sigma ^{ - 1}}\mathtt{vec}\left[ {{\partial _\varphi }\sigma } \right]}\\
	{\mathtt{vec}{{\left[ {{\partial _\varphi }\sigma } \right]}^\dag }{\Sigma ^{ - 1}}\mathtt{vec}\left[ {{\partial _r}\sigma } \right]}&{\mathtt{vec}{{\left[ {{\partial _\varphi }\sigma } \right]}^\dag }{\Sigma ^{ - 1}}\mathtt{vec}\left[ {{\partial _\varphi }\sigma } \right]}
	\end{array}} \right].
\end{equation}}
Using the identity ${\left( {A \otimes B} \right)^{ - 1}} = {A^{ - 1}} \otimes {B^{ - 1}}$, we obtain the RLD quantum Fisher information matrix as
{\small \begin{equation}
	\mathcal{F} =\left[ {\begin{array}{*{20}{c}}
		{\frac{{{{\left( {1 + 2 \bar n} \right)}^2}\left( {1 + 2 \bar n\left( {1 +\bar n} \right)} \right)}}{{2{\bar n^2}{{\left( {1 +\bar n} \right)}^2}}}}&{\frac{{{\rm{i}}{{\left( {1 + 2 \bar n} \right)}^3}\sinh \left[ {2r} \right]}}{{2{ \bar n^2}{{\left( {1 + \bar n} \right)}^2}}}}\\
		{ - \frac{{{\rm{i}}{{\left( {1 + 2 \bar n} \right)}^3}\sinh \left[ {2r} \right]}}{{2{\bar n^2}{{\left( {1 +\bar n} \right)}^2}}}}&{\frac{{\left( {1 + 2 \bar n\left( {1 +\bar n} \right)} \right)\sinh {{\left[ {2r} \right]}^2}}}{{2{\bar n^2}{{\left( {1 +\bar n} \right)}^2}{{\left( {1 + 2 \bar n} \right)}^{ - 2}}}}}
		\end{array}} \right].
	\end{equation}}
Similarly, it is easy to verify that $\det \mathcal{M} \ne 0$, (i.e. $\mathcal{M}$ is invertible) and the SLD quantum Fisher information matrix (\ref{38}) is given by 
	\begin{equation}
H = \left[ {\begin{array}{*{20}{c}}
	{\frac{{4 \bar n\left( {\bar n + 1} \right) + 1}}{{\bar n\left( {\bar n + 1} \right)}}}&0\\
	0&{\frac{{{{\left( {1 + 2\bar n} \right)}^2}\sinh {{\left[ {2r} \right]}^2}}}{{\bar n\left( {1 + \bar n} \right)}}}
	\end{array}} \right]. \label{61}
	\end{equation}
The two quantum Cramér-Rao bounds can be simply evaluated from Eqs. (\ref{21}) and (\ref{22}). They are given by 
\begin{equation}
{B_R} = \frac{{\left( {1 + 2\bar n\left( {1 + \bar n} \right)} \right)\coth {{\left[ {2r} \right]}^2}\sinh \left[ {2r} \right] + 2\left( {1 + 2 \bar n} \right)}}{{2{{\left( {1 + 2 \bar n} \right)}^2}\sinh \left[ {2r} \right]}}, \label{62}
\end{equation}
\begin{equation}
{B_S} = \frac{{\bar n\left( {1 + \bar n} \right)\coth {{\left[ {2r} \right]}^2}}}{{{{\left( {1 + 2\bar n} \right)}^2}}}. \label{63}
\end{equation}
The expressions (\ref{62}) and (\ref{63}) show that the precision of the estimated parameters does not depend on the value of the unknown phase rotation parameter, but depends only on the squeezing parameter and the inverse of the temperature.
The SLD quantum Fisher information matrix (\ref{61}) is diagonal, this diagonal form follows from  ${\rm{Im}}\left( {Tr\left[ {{{\rho }_{out}}\hat L_{{\theta _\mu }}^S\hat L_{{\theta _\nu }}^S} \right]} \right) = 0$  which means that the quantum Cramér-Rao bound is attainable.

\begin{figure}[h]
	\centering
	\includegraphics[width=8.2cm]{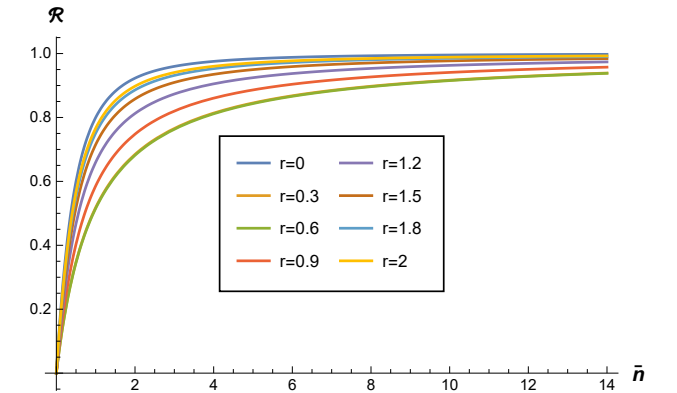}
	\captionsetup{justification=raggedright, singlelinecheck=false}\captionof{figure}{The plot of the ratio between the symmetric logarithmic derivative (SLD) and the right logarithmic derivative (RLD) quantum Cramér-Rao bound ($\mathcal{R}$) as a function of the thermal mean photon number for the various values of the squeezing parameter.}\label{F1}
\end{figure}
The behavior represented in Fig. (\ref{F1}) shows that the ratio $\mathcal{R}$ is less than 1 when $\bar n$ is small and then increases when $r$ and $\bar n$ increase until to attain the value 1 for large values of the mean photons number. Therefore, we conclude that $B_{MI} = B_S = B_R$ when the thermal mean photon number takes the large values, and $B_{MI} = B_R$ otherwise. It is also possible to analyze these results in terms of the temperature which characterized the thermal state. In fact, this can be simply done by performing the  substitution $2\bar n + 1 = \coth \left( {{\raise0.7ex\hbox{$\beta $} \!\mathord{\left/{\vphantom {\beta  2}}\right.\kern-\nulldelimiterspace} \!\lower0.7ex\hbox{$2$}}} \right)$.

\begin{figure}[h]
	\centering
	\includegraphics[width=8.2cm]{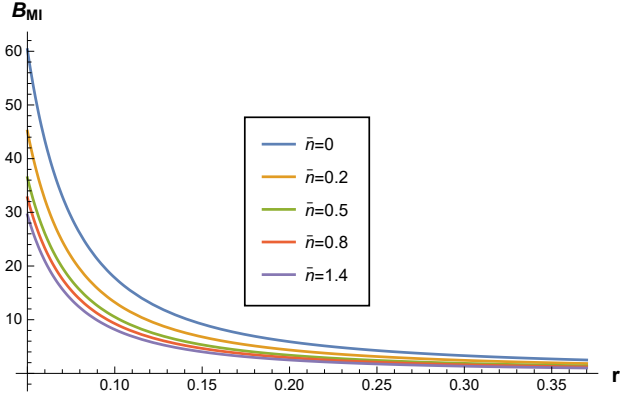}
	\captionsetup{justification=raggedright, singlelinecheck=false}\captionof{figure}{The plot of the most informative quantum Cramér-Rao bound $B_{MI}$ for the joint estimation of squeezing parameter $r$ and phase rotation $\varphi$ as a function of $r$ for the various values of thermal mean photon number $\bar n$ in the probe thermal state. }\label{F2}
\end{figure}
The results represented in Fig. (\ref{F2})  shows that $B_{MI}$ decreases with increasing values of the mean energy of the probe thermal state $\bar n$, while it reaches their minimum value when $r$ takes great values. This means that the optimal values of simultaneous estimation of the parameters $r$ and $\varphi$ is achievable when we increase the mean energy of the probe thermal state.

Now, we consider a new estimation problem with the coherent state $\left| \alpha  \right\rangle $ as input state evolving in the same  channel Gaussian described by the transformation $\hat R\left( \varphi  \right)\hat S\left( r \right)$, such as
\begin{equation}
{\rho _{out}} = \hat R\left( \varphi  \right)\hat S\left( r \right)\left| \alpha  \right\rangle \left\langle \alpha  \right|{\hat S^\dag }\left( r \right){\hat R^\dag }\left( \varphi  \right). \label{64}
\end{equation}
The input coherent state $\left| \alpha  \right\rangle $ is characterized by
\begin{equation}
{\mathbf{d}_{in}} =2\left[ {\begin{array}{*{20}{c}}
	{{\mathop{\rm Re}\nolimits} \left[ \alpha  \right]}\\
	{{\mathop{\rm Im}\nolimits} \left[ \alpha  \right]}
	\end{array}} \right], \hspace{1cm} {\sigma _{in}} = \left[ {\begin{array}{*{20}{c}}
	1&0\\
	0&1
	\end{array}} \right].
\end{equation}
Using the symplectic transformations (\ref{53}), we can express the first and second moments of the output state (\ref{64}) as follows
 \begin{equation}
 {\mathbf{d}_{out}} = \hat R\left( \varphi  \right)\hat S\left( r \right){\mathbf{d}_{in}}, \hspace{0.3cm}{\sigma _{out}} = \hat R\left( \varphi  \right)\hat S\left( r \right){\sigma _{in}}{{\hat S}^\dag }\left( r \right){{\hat R}^\dag }\left( \varphi  \right).
 \end{equation}
To calculate the RLD and SLD quantum Fisher information matrix, we first derive
\begin{equation} 
\mathtt{vec}\left[ {{\partial _r}{\sigma _{out}}} \right] = 2\left[ {\begin{array}{*{20}{c}}
	{{{\sin }^2}\varphi {e^{2r}} - {{\cos }^2}\varphi {e^{ - 2r}}}\\
	{2\sin \varphi \cos \varphi \sinh 2r}\\
	{2\sin \varphi \cos \varphi \sinh 2r}\\
	{{{\cos }^2}\varphi {e^{2r}} - {{\sin }^2}\varphi {e^{ - 2r}}}
	\end{array}} \right],
\end{equation}
\begin{equation}
\mathtt{vec}\left[ {{\partial _\varphi }{\sigma _{out}}} \right] = 2\sinh 2r\left[ {\begin{array}{*{20}{c}}
	{2\sin \varphi \cos \varphi }\\
	{\cos 2\varphi }\\
	{\cos 2\varphi }\\
	{ - 2\sin \varphi \cos \varphi }
	\end{array}} \right],
\end{equation}
\begin{equation}
{\partial _r}{\mathbf{d}_{out}} = 2\left[ {\begin{array}{*{20}{c}}
	{ - {{\rm{e}}^{ - r}}{\mathop{\rm Re}\nolimits} \left[ \alpha  \right]\cos \varphi  + {{\rm{e}}^r}{\mathop{\rm Im}\nolimits} \left[ \alpha  \right]\sin \varphi }\\
	{{{\rm{e}}^r}{\mathop{\rm Im}\nolimits} \left[ \alpha  \right]\cos \varphi  + {{\rm{e}}^{ - r}}{\mathop{\rm Re}\nolimits} \left[ \alpha  \right]\sin \varphi }
	\end{array}} \right],
\end{equation}
\begin{equation}
{\partial _\varphi }{\mathbf{d}_{out}} = 2\left[ {\begin{array}{*{20}{c}}
	{{{\rm{e}}^r}{\mathop{\rm Re}\nolimits} \left[ \alpha  \right]\cos \varphi  - {{\rm{e}}^{ - r}}{\mathop{\rm Im}\nolimits} \left[ \alpha  \right]\sin \varphi }\\
	{ - {{\rm{e}}^{ - r}}{\mathop{\rm Im}\nolimits} \left[ \alpha  \right]\cos \varphi  - {{\rm{e}}^r}{\mathop{\rm Re}\nolimits} \left[ \alpha  \right]\sin \varphi }
	\end{array}} \right].
\end{equation}
Obviously, $ \det \Gamma  = 0$ ($\Gamma$ is singular matrix) due to the saturation of principle uncertainty given by Eq. (\ref{10}) for coherent states. We use the Tikhonov regularization to compute the Moore-Penrose pseudoinverse of $\Gamma$. This gives 
{\small  \begin{equation}
	 {\Gamma ^ + } = {{\rm{e}}^{2r}}\left[ {\begin{array}{*{20}{c}}
	 	{\frac{{{\lambda _ + }\left( r \right) - {\lambda _ - }\left( r \right)\cos \left[ {2\varphi } \right]}}{{2{\lambda _ + }{{\left( r \right)}^2}}}}&{\frac{{2{\rm{i}}{{\rm{e}}^{2r}} + {\lambda _ - }\left( r \right)\sin \left[ {2\varphi } \right]}}{{2{\lambda _ + }{{\left( r \right)}^2}}}}\\
	 	{\frac{{ - 2{\rm{i}}{{\rm{e}}^{2r}} + {\lambda _ - }\left( r \right)\sin \left[ {2\varphi } \right]}}{{2{\lambda _ + }{{\left( r \right)}^2}}}}&{\frac{{{\lambda _ + }\left( r \right) + {\lambda _ - }\left( r \right)\cos \left[ {2\varphi } \right]}}{{2{\lambda _ + }{{\left( r \right)}^2}}}}
	 	\end{array}} \right], 
	 \end{equation}}
 where ${\lambda _ \pm }\left( r \right) = {{\rm{e}}^{4r}} \pm 1$. We note that ${\Gamma ^\dag } = \Gamma $. Using ${\Sigma ^ + } = {\Gamma ^ + } \otimes {\Gamma ^ + }$ the RLD quantum Fisher information matrix is obtained from (\ref{31}), as
 
\begin{widetext}
	{\small \begin{equation}
\mathcal{F} = \left[ {\begin{array}{*{20}{c}}
	{\frac{{2\left( {{{\left( {1 + {{\rm{e}}^{8r}}} \right)}^2} + 4{\lambda _ + }{{\left( r \right)}^2}\left( {{\mathop{\rm Re}\nolimits} {{\left[ \alpha  \right]}^2} + {{\rm{e}}^{8r}}{\mathop{\rm Im}\nolimits} {{\left[ \alpha  \right]}^2}} \right)} \right)}}{{{\lambda _ + }{{\left( r \right)}^4}}}}&{\frac{{2{{\rm{e}}^{2r}}\left( { - {\rm{i}}\left( {{\lambda _ - }\left( r \right)\left( {1 + {{\rm{e}}^{8r}}} \right)} \right) - 2{\lambda _ + }{{\left( r \right)}^2}\alpha \left( { - {\rm{i}}{\mathop{\rm Re}\nolimits} \left[ \alpha  \right] + {{\rm{e}}^{4r}}{\mathop{\rm Im}\nolimits} \left[ \alpha  \right]} \right)} \right)}}{{{\lambda _ + }{{\left( r \right)}^4}}}}\\
	{\frac{{2{{\rm{e}}^{2r}}\left( {{\rm{i}}\left( {{\lambda _ - }\left( r \right)\left( {1 + {{\rm{e}}^{8r}}} \right)} \right) - 2{\lambda _ + }{{\left( r \right)}^2}{\alpha ^*}\left( {{\rm{i}}{\mathop{\rm Re}\nolimits} \left[ \alpha  \right] + {{\rm{e}}^{4r}}{\mathop{\rm Im}\nolimits} \left[ \alpha  \right]} \right)} \right)}}{{{\lambda _ + }{{\left( r \right)}^4}}}}&{\frac{{2{{\rm{e}}^{4r}}\left( {{\lambda _ - }{{\left( r \right)}^2} + {\lambda _ + }{{\left( r \right)}^2}{{\left| \alpha  \right|}^2}} \right)}}{{{\lambda _ + }{{\left( r \right)}^4}}}}
	\end{array}} \right].
	\end{equation}}
\end{widetext}
Similarly, we can calculate the Moore-Penrose pseudoinverse of $\mathcal{M}$ by mean of Tikhonov regularization. In this picture using (\ref{37}),  we obtain the following SLD quantum Fisher information matrix   
{\small\begin{equation}
 H = \left[ {\begin{array}{*{20}{c}}
 	{2\left( {4{{\left| \alpha  \right|}^2} + \tanh {{\left[ {4r} \right]}^2}} \right)}&{ - 16{\mathop{\rm Re}\nolimits} \left[ \alpha  \right]{\mathop{\rm Im}\nolimits} \left[ \alpha  \right]\cosh \left[ {2r} \right]}\\
 	{ - 16{\mathop{\rm Re}\nolimits} \left[ \alpha  \right]{\mathop{\rm Im}\nolimits} \left[ \alpha  \right]\cosh \left[ {2r} \right]}&{{\rm{8}}{{\rm{e}}^{ - 4r}}\left( {{{\rm{e}}^{8r}}{\mathop{\rm Re}\nolimits} {{\left[ \alpha  \right]}^2} + {\mathop{\rm Im}\nolimits} {{\left[ \alpha  \right]}^2}} \right)}
 	\end{array}} \right].
\end{equation}}
The two bounds $B_R$  and $B_S$  for this protocol can be computed from Eqs. (\ref{21}) and (\ref{22}). One gets
	{\small \begin{equation}
	{B_R} = \frac{{f\left( {r,\alpha } \right)\cosh {{\left[ {2r} \right]}^2} + g\left( {r,\alpha } \right)\cosh {{\left[ {2r} \right]}^3} + h\left( {r,\alpha } \right)\sinh \left[ {2r} \right]}}{{2\left( {{\mathop{\rm Im}\nolimits} {{\left[ \alpha  \right]}^2} + {{\rm{e}}^{8r}}{\mathop{\rm Re}\nolimits} {{\left[ \alpha  \right]}^2}} \right)}},\label{75}
\end{equation}}
{\small \begin{equation}
{B_S} = \frac{{{\rm{4}}k\left( {r,\alpha } \right) + {{\rm{e}}^{4r}}\left( {4{{\left| \alpha  \right|}^2} + \tanh {{\left[ {4r} \right]}^2}} \right)}}{{8\left( {4{{\left( {{\mathop{\rm Re}\nolimits} {{\left[ \alpha  \right]}^2} - {{\rm{e}}^{4r}}{\mathop{\rm Im}\nolimits} {{\left[ \alpha  \right]}^2}} \right)}^2} + k\left( {r,\alpha } \right)\tanh {{\left[ {4r} \right]}^2}} \right)}},\label{76}
\end{equation}}
where $f\left( {r,\alpha } \right) = 1 + {{\rm{e}}^{8r}}\left( {1 + 4{\mathop{\rm Im}\nolimits} {{\left[ \alpha  \right]}^2}} \right) + 4{\mathop{\rm Re}\nolimits} {\left[ \alpha  \right]^2} + {{\rm{e}}^{4r}}\left( {4{{\left| \alpha  \right|}^2} - 1} \right)$, $g\left( {r,\alpha } \right) = 8{{\rm{e}}^{4r}}\left( {{\mathop{\rm Im}\nolimits} {{\left[ \alpha  \right]}^2} - {\mathop{\rm Re}\nolimits} {{\left[ \alpha  \right]}^2}} \right)$, $h\left( {r,\alpha } \right) = 2{{\rm{e}}^{4r}}\left( {2{{\left| \alpha  \right|}^2} + \left( {2 + 4{{\left| \alpha  \right|}^2}} \right)\cosh \left[ {4r} \right]} \right)$, and $k\left( {r,\alpha } \right) = {{\rm{e}}^{8r}}{\mathop{\rm Im}\nolimits} {\left[ \alpha  \right]^2} + {\mathop{\rm Re}\nolimits} {\left[ \alpha  \right]^2}$. Eqs. (\ref{75}) and (\ref{76}) show that RLD and SLD quantum Cramér-Rao bounds depend on the squeezing parameter $r$ and the variable labeling the coherent state (the input state). Here also, it is interesting to note that the rotation parameter $\varphi$ does not contribute to the quantum Cramér-Rao bounds.

\begin{figure}[t]
	\centering
	\includegraphics[width=8.2cm]{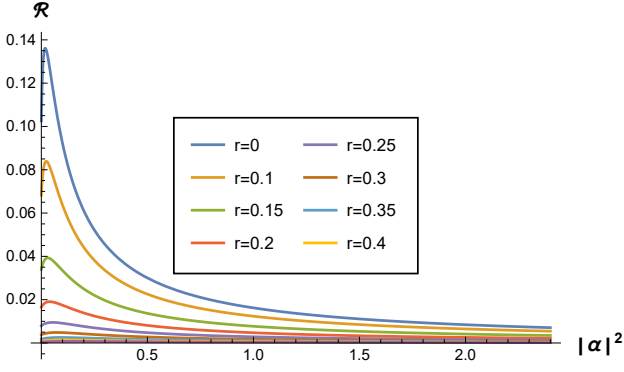}
	\captionsetup{justification=raggedright, singlelinecheck=false}\captionof{figure}{The plot of the ratio between the symmetric logarithmic derivative (SLD) and the right logarithmic derivative (RLD) quantum Cramér-Rao bound ($\mathcal{R}$) as a function of the mean photon number of a coherent state for the various values of the squeezing parameter.}\label{F3}
\end{figure}

 From the results represented in Fig. (\ref{F3}), we notice that the ratio $\mathcal{R}$ is always less than 1. Consequently, the most informative quantum Cramér-Rao bound corresponds to RLD quantum Cramér-Rao bound ($B_{MI}$=$B_R$). This result can be explained by the fact that the condition of saturation of the SLD quantum CR bound is not satisfied. Using the Eq .(\ref{45}), we derive
\begin{equation}
{\rm{Im}}\left( {Tr\left[ {{\rho _{out}}\hat L_{{\theta _\mu }}^S\hat L_{{\theta _\nu }}^S} \right]} \right) = 8\left( {{{\rm{e}}^{ - 2r}}{\mathop{\rm Re}\nolimits} {{\left[ \alpha  \right]}^2} - {{\rm{e}}^{2r}}{\mathop{\rm Im}\nolimits} {{\left[ \alpha  \right]}^2}} \right).
\end{equation}
This quantity is not zero, which means that the SLD quantum Cramér-Rao bound can not be saturated for simultaneous estimation of $r$ and $\varphi$ encoded into squeezing and rotation operators respectively when taking the coherent state as input state.

\begin{figure}[t]
	\centering
	\includegraphics[width=8.2cm]{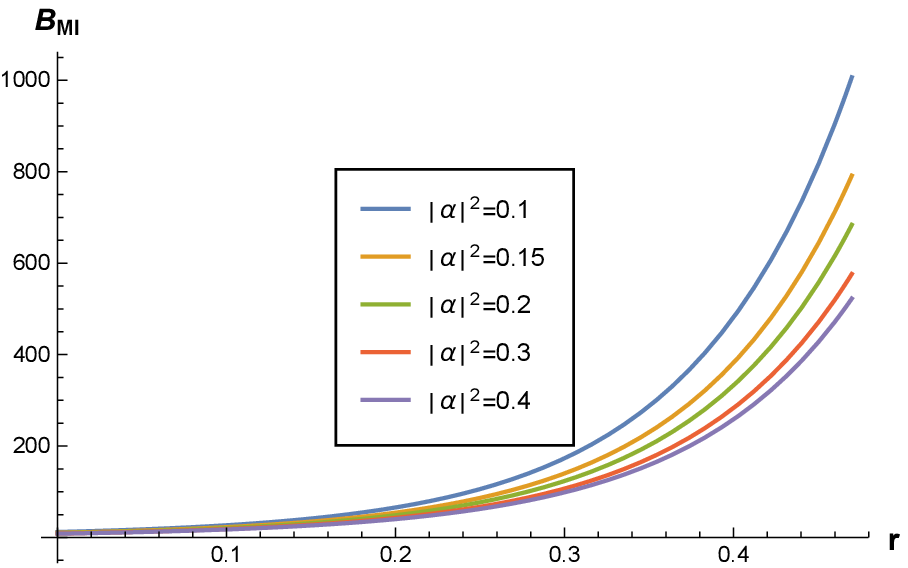}
	\captionsetup{justification=raggedright, singlelinecheck=false}\captionof{figure}{The plot of the most informative quantum Cramér-Rao bound $B_{MI}$ for the joint estimation of squeezing parameter $r$ and phase rotation $\varphi$ as a function of $r$ for the various values of the mean photon number of a coherent state ${\left| \alpha  \right|^2}$.}\label{F4}
\end{figure}
The behavior represented in Fig. (\ref{F4}) shows that $B_{MI}$ increases when the energy of the probe coherent states $\vert \alpha\vert^2$ decreases.  It reaches their minimum value when $r$ takes smaller values, which implies that the optimal values for simultaneous estimation of parameters $r$ and $\varphi$ is achievable when we increase the mean energy of the probe coherent state.

From the comparison between the results represented in Fig. (\ref{F2}) and the results represented in  Fig. (\ref{F4}), we note that there are similarities between the performance obtained for the simultaneous estimation of $r$ and $\varphi$ encoded respectively in squeezing and rotation operators  when taking the thermal state and the coherent state as probes states. But it must noticed that thermal states present more advantages when the mean energy of coherent and thermal states takes the smaller values.
\section{conclusion}

Quantum Cramér-Rao bound is the key tool used to estimate unknown parameters in a quantum system. This bound is determined from the quantum Fisher information matrix. In this article, we determined the expressions of the RLD quantum Fisher information matrix and SLD quantum Fisher information matrix by explicitly computing the expressions of the right logarithmic derivative (RLD) and symmetric logarithmic derivative (SLD) operators corresponding for the multi-mode quantum Gaussian states. We also expressed the saturation condition of quantum Cramér-Rao bounds associated with SLD operator in multiparameter quantum estimation protocols. We then illustrated the derived formalism with some examples of quantum Gaussian channels.

We note that all the explicit expressions of the quantities supplying this general formalism are expressed in terms of the first and second moments. That brings back to the degree of freedom of the quantum Gaussian states which limits only to two characteristic parameters (first and second moments). This remarkable advantage is an incentive to provide more general strategies in estimation multiparameter quantum Gaussian metrology.
\begin{widetext}
	\appendix
	\section{Right Logarithmic Derivative (RLD)  \label{app:A}}
	To determine the expressions of RLD and SLD quantum Fisher information matrix and the corresponding logarithmic derivatives operators, we need to use certain properties of the characteristic function of Gaussian states which established a link between the infinite-dimensional Hilbert space and finite-dimensional phase-space. The characteristic function is defined by
	\begin{equation}
	{\chi _{\hat \rho }} = Tr\left[ {\hat D\hat \rho } \right] \mathop { =\joinrel=}\limits^{(\ref{6})} Tr\left[ {{e^{i{{\mathbf{\tilde q}}^\mathtt{T}}\mathbf{\hat q}}}{e^{i{{\mathbf{\tilde p}}^\mathtt{T}}\mathbf{\hat p}}}{e^{\frac{i}{2}{{\mathbf{\tilde q}}^\mathtt{T}}\mathbf{\tilde p}}}\hat \rho } \right] = Tr\left[ {{e^{i{{\mathbf{\tilde p}}^\mathtt{T}}\mathbf{\hat p}}}{e^{i{{\mathbf{\tilde q}}^\mathtt{T}}\mathbf{\hat q}}}{e^{ - \frac{i}{2}{{\mathbf{\tilde q}}^\mathtt{T}}\mathbf{\tilde p}}}\hat \rho } \right], \label{A1}
	\end{equation}
	where $\mathbf{\hat q} (\mathbf{\hat p})$ and $\mathbf{\tilde q} (\mathbf{\tilde p})$ are the vectors of odd (even) entries of the parent vectors $\mathbf{\hat r}$ and $\mathbf{\tilde r}$ respectively. This decomposition follows from the Baker-Campbell-Hausdorff formula.
	
	The derivation of (\ref{A1}) with respect to ${{\tilde r}_k}({{\tilde q}_k},{{\tilde p}_k})$ gives the following interesting identities;
	\begin{equation}
	Tr\left[ {\hat D \hspace{0.1cm}\hat \rho \hspace{0.1cm}{{\hat r}_k}} \right] = \left( { - i{\mkern 1mu} {\partial _{{{\tilde r}_{_k}}}} - \frac{1}{2}{\Omega _{kk'}}{{\tilde r}_{k'}}} \right){\chi _{\hat \rho }},  \hspace*{1cm} Tr\left[ {\hat D \hspace{0.1cm} \hat \rho \hspace{0.1cm}  {{\hat r}_j} \hspace{0.1cm} {{\hat r}_k}} \right] = \left( { - i{\mkern 1mu} {\partial _{{{\tilde r}_k}}} - \frac{1}{2}{\Omega _{kk'}}{{\tilde r}_{k'}}} \right)\left( { - i{\mkern 1mu} {\partial _{{{\tilde r}_{_j}}}} - \frac{1}{2}{\Omega _{jj'}}{{\tilde r}_{j'}}} \right){\chi _{\hat \rho }}.\label{A2}
	\end{equation}
	\begin{equation}
	Tr\left[ {\hat D \hspace{0.1cm}\left( {\hat \rho \hspace{0.1cm}{{\hat r}_j} + {{\hat r}_j} \hspace{0.1cm}\hat \rho } \right)} \right] =  - 2i{\mkern 1mu} {\partial _{{{\tilde r}_j}}}{\chi _{\hat \rho }}, \hspace{2cm} Tr\left[ {\hat D \hspace{0.1cm}\left( {\hat \rho \hspace{0.1cm} {{\hat r}_j}\hspace{0.1cm}{{\hat r}_k} + {{\hat r}_k}\hspace{0.1cm}{{\hat r}_j}\hspace{0.1cm}\hat \rho } \right)} \right] = \frac{1}{2}\left( {{\Omega _{jj'}}{{\tilde r}_{j'}}{\Omega _{kk'}}{{\tilde r}_{k'}} - 4{\partial _{{{\tilde r}_k}}}{\partial _{{{\tilde r}_j}}}} \right){\chi _{\hat \rho }}.\label{A3}
	\end{equation}
We derive the expression of quantum Gaussian states (\ref{7}), which is expressed in terms of the first and the second moments with respect to  ${{\tilde r}_l}$ and ${\theta _\mu }$(estimated parameters) respectively. We get 
	\begin{equation}
	{\partial _{{\theta _\mu }}}{\chi _{\hat \rho }} = \left( {i{\mkern 1mu} {{\tilde r}_v}{\partial _{{\theta _\mu }}}{d_v} - \frac{1}{4}{\partial _{{\theta _\mu }}}{\sigma _{pm}}{{\tilde r}_p}{{\tilde r}_m}} \right){\chi _{\hat \rho }}, \hspace*{0.8cm} 	{\partial _{{{ {\tilde r}}_l}}}{\chi _{\hat \rho }} = \left( {i{\mkern 1mu} {\partial _{{\theta _\mu }}}{d_l} - \frac{1}{2}{\partial _{{\theta _\mu }}}{\sigma _{lm}}{{\tilde r}_l}} \right){\chi _{\hat \rho }}. \label{A4}
	\end{equation}
These expressions are important in computing the explicit expression of RLD (\ref{27}). For this one needs to calculate the expressions of  ${{\cal L}^R}^{\left( 0 \right)}$, ${\cal L}^{R\left( 1 \right)}$ and ${\cal L}^{R\left( 2 \right)}$ occurring in Eq. (\ref{27}). We refer each passage by the formula used
	\begin{align}
	{\partial _{{\theta _\mu }}}{\mathbf{\chi} _{\hat \rho }}&= Tr\left[ {\hat D\hspace{0.1cm}{\partial _{{\theta _\mu }}}\hat \rho } \right] \\ \notag&
\mathop{ =\joinrel=}\limits^{\left( \ref{13} \right)} Tr\left[ {\hat D\hspace{0.1cm}\hat \rho\hspace{0.1cm} \mathcal{\hat L_{{\theta _\mu }}}^R} \right]\\ \notag&
	\mathop  { =\joinrel=} \limits^{(\ref{27})} {\mathcal{\mathcal{L}}^R}^{\left( 0 \right)}Tr\left[ {\hat D\hspace{0.1cm}\hat \rho } \right] + \mathcal{\mathcal{L}}_l^{R\left( 1 \right)}Tr\left[ {\hat D\hspace{0.1cm}\hat \rho \hspace{0.1cm}{{\hat r}_l}} \right] + \mathcal{L}_{jk}^{R\left( 2 \right)}Tr\left[ {\hat D \hspace{0.1cm}\hat \rho\hspace{0.1cm} {{\hat r}_j}\hspace{0.1cm}{{\hat r}_k}} \right]\\ \notag &
	\mathop { =\joinrel=}\limits^{(\ref{A2})} {{\cal L}^R}^{\left( 0 \right)}{{\bf{\chi }}_{\hat \rho }} + {\cal L}_l^{R\left( 1 \right)}\left( { - i{\partial _{{{\tilde r}_l}}} - \frac{1}{2}{\Omega _{ll'}}{{\tilde r r}_{l'}}} \right){\chi _{\hat \rho }} + {\cal L}_{jk}^{R\left( 2 \right)}\left( { - i{\partial _{{\tilde r_k}}} - \frac{1}{2}{\Omega _{kk'}}{{\bar r}_{k'}}} \right)\left( { - i{\partial _{{{\tilde r}_j}}} - \frac{1}{2}{\Omega _{jj'}}{{\tilde r}_{j'}}} \right){\chi _{\hat \rho }}. \label{A6}
	\end{align}
	Using the results of Eq.(\ref{A4}), one finds
	\begin{align}
	\left( {i{{\tilde r}_v}{\partial _{{\theta _\mu }}}{d_v} - \frac{1}{4}{\partial _{{\theta _\mu }}}{\sigma _{mp}}{{\tilde r}_m}{{\tilde r}_p}} \right){\chi _{\hat \rho }} = &{{\cal L}^R}^{\left( 0 \right)}{\chi _{\hat \rho }} + {\cal L}_l^{R\left( 1 \right)}\left( {\frac{i}{2}{\sigma _{ll'}}{{\tilde r}_{l'}} + {d_l} - \frac{1}{2}{\Omega _{i{i^,}}}{{\tilde r}_{{i^,}}}} \right){\chi _{\hat \rho }} + \\ \notag&  {\cal L}_{jk}^{R(2)}\left( {\left( {\frac{1}{2}{\sigma _{jj'}}{{\tilde r}_{j'}} - i{d_j}} \right)\left( { - \frac{1}{2}{\sigma _{kk'}}{{\tilde r}_{k'}} + i{d_k}} \right) + \frac{1}{2}{\sigma _{jk}}+{\frac{i}{2}{\Omega _{jk}}}} \right){\chi _{\hat \rho }} - \\ \notag& {\cal L}_{jk}^{R(2)}\left( {\frac{i}{4}{\Omega _{jj'}}{\sigma _{kk'}}{{\tilde r}_{j'}}{{\tilde r}_{k'}} + \frac{i}{4}{\Omega _{kk'}}{\sigma _{jj'}}{{\tilde r}_{k'}}{{\tilde r}_{j'}} + \frac{1}{2}{\Omega _{jj'}}{{\tilde r}_{j'}}{d_k} + \frac{1}{2}{\Omega _{kk'}}{{\tilde r}_{k'}}{d_j} - \frac{1}{4}{\Omega _{jj'}}{\Omega _{kk'}}{{\tilde r}_{j'}}{{\tilde r}_{k'}}} \right){\chi _{\hat \rho }}. 
	\end{align}
	Now, we can amend the different orders of the last equation independently. We note that ${\chi _{\hat \rho }}$ is always non zero;

		The identification of second-order terms of Eq. (\ref{A6}) leads to:
	
	\begin{equation}
	- {\partial _{{\theta _\mu }}}{\sigma _{mp}}{\tilde r_m}{{\tilde r}_p} = {\cal L}_{jk}^{R(2)}\left( {{\Omega _{jj'}}{{\tilde r}_{j'}}{\Omega _{kk'}}{{\tilde r}_{k'}} - {\sigma _{jj'}}{{\tilde r}_{j'}}{\sigma _{kk'}}{{\tilde r}_{k'}} - i\,{\sigma _{kk'}}{{\tilde r}_{k'}}{\Omega _{jj'}}{{\tilde r}_{j'}} - i\,{\Omega _{kk'}}{{\tilde r}_{k'}}{\sigma _{jj'}}{{\tilde r}_{j'}}} \right).
	\end{equation}
	Using a matrix representation (without index), one finds
	\begin{align}
	{\partial _{{\theta _\mu }}}\sigma  &= \sigma {{\cal L}^{\left( R \right)}}^{\left( 2 \right)}\sigma  - \Omega {{\cal L}^{\left( R \right)}}^{\left( 2 \right)}\Omega  + i\,\sigma {{\cal L}^{\left( R \right)}}^{\left( 2 \right)}\Omega  + i\,\Omega {{\cal L}^{\left( R \right)}}^{\left( 2 \right)}\sigma \\ \notag&
	= {\Gamma}{{\cal L}^{\left( R \right)}}^{\left( 2 \right)}{\Gamma },
	\end{align}
	where ${\Gamma} = \sigma  + i\,\Omega $. To determine the expression of ${{\cal L}^{\left( R \right)}}^{\left( 2 \right)}$, one employs the property
	\begin{equation}
	\mathtt{vec}\left[ {ABC} \right] = \left( {{{C^\dag }} \otimes A} \right)\mathtt{vec}\left[ B \right], \label{A10}
	\end{equation}
	where $A$, $B$, and $C$ are matrices with complex elements. Thus one gets
	\begin{equation}
	\mathtt{vec}\left[ {{\mathcal{L}^{\left( R \right)}}^{\left( 2 \right)}} \right] = {\left( {\Gamma ^\dag \otimes {\Gamma }} \right)^{+}}\mathtt{vec}\left[ {{\partial _{{\theta _\mu }}}\sigma } \right]. \label{A11}
	\end{equation}
The identification of first-order terms of Eq. (\ref{A6}) leads to:
	\begin{align}
	i{{\tilde r}_v}{\partial _{{\theta _\mu }}}{d_v} = \mathcal{L}_l^{R\left( 1 \right)}\left( {\frac{i}{2}{\sigma _{ll'}}{{\tilde r}_{l'}} - \frac{1}{2}{\Omega _{i{i^,}}}{{\tilde r}_{{i^,}}}} \right){\chi _{\hat \rho }} + {\rm{ }}\mathcal{L}_{jk}^{R(2)}\left( {\frac{i}{2}{\sigma _{jj'}}{{\tilde r}_{j'}}{d_k} + \frac{i}{2}{\sigma _{kk'}}{{\tilde r}_{k'}}{d_j} - \frac{1}{2}{\Omega _{jj'}}{{\tilde r}_{j'}}{d_k} - \frac{1}{2}{\Omega _{kk'}}{{\tilde r}_{k'}}{d_j}} \right).
	\end{align}
The matrix form of the last equation writes
	\begin{align}
	{\partial _{{\theta _\mu }}}\mathbf{d} =\frac{1}{2}\left( {\sigma  + i\,\Omega } \right){\mathcal{L}^R}^{\left( 1 \right)} + {\rm{ }}\left( {\sigma  + i\,\Omega } \right){\mathcal{L}^R}^{\left( 2 \right)}\mathbf{d}
	= \frac{1}{2}{\Gamma}{\mathcal{L}^R}^{\left( 1 \right)} + {\rm{ }}{\Gamma}{\mathcal{L}^R}^{\left( 2 \right)}\mathbf{d},
	\end{align}
and the expression of ${\mathcal{L}^R}^{\left( 1 \right)}$ is found as
	\begin{equation}
	{\mathcal{L}^R}^{\left( 1 \right)} = 2\Gamma ^{+}{\partial _{{\theta _\mu }}}\mathbf{d}{\rm{  - 2}}{\mathcal{L}^R}^{\left( 2 \right)}\mathbf{d}.
	\end{equation}
The identification of zero-order terms of Eq. (\ref{A6}) leads to:
	\begin{equation}
	0 = {\mathcal{L}^R}^{\left( 0 \right)} +\mathcal{ L}_l^{R\left( 1 \right)}{d_l} + \mathcal{L}_{jk}^{R(2)}\left( {{d_j}{d_k} + \frac{1}{2}{\sigma _{jk}} + \frac{i}{2}{\Omega _{jk}}} \right),
	\end{equation}
which takes the following matrix form
	\begin{equation}
	0 = {{\cal L}^R}^{\left( 0 \right)} + {{\cal L}^{R\left( 1 \right)}}^T{\bf{d}} + {{\bf{d}}^T}{{\cal L}^{R\left( 2 \right)}}{\bf{d}} + \frac{1}{2}Tr\left[ {{\Gamma }{{\cal L}^{R\left( 2 \right)}}} \right].
	\end{equation}
	We find that the expression of ${\mathcal{L}^R}^{\left( 0 \right)}$ is given by
	\begin{equation}
	{\mathcal{L}^R}^{\left( 0 \right)} =  - \frac{1}{2}Tr\left[ {{\Gamma }{\mathcal{L}^{R\left( 2 \right)}}} \right] - {\mathcal{L}^{R\left( 1 \right)}}^T \mathbf{d} - {\mathbf{d}^T}{\mathcal{L}^{R\left( 2 \right)}}\mathbf{d}.
	\end{equation}
	
	\section{RLD Quantum Fisher Information Matrix (QFIM) \label{App:B}}
Inserting the expression of the right logarithmic derivative (RLD) obtained in the Appendix (\ref{app:A}) into the definition of the RLD quantum Fisher information matrix (\ref{15}) which writes also as
	\begin{equation}
		{{\cal F}_{{\theta _\mu }{\theta _\nu }}} =Tr\left[ {{\partial _{{\theta _\mu }}}\hat \rho {\cal L}{{_{{\theta _\nu }}^R}^\dag }} \right], \label{B1}
	\end{equation}
and using the property of the characteristic function if it is evaluated when $\mathbf{\tilde r}=0$
	\begin{equation}
	Tr\left[ {\hat \rho } \right] = {\left. {Tr\left[ {\hat D\hat \rho } \right]} \right|_{\mathbf{\tilde r} = 0}} = {\left. {{\chi _{\hat \rho }}} \right|_{\mathbf{\tilde r} = 0}} = 1, \label{B2}
	\end{equation}
the expression of (\ref{B1})  can be written as
	\begin{align}
	{{\cal F}_{{\theta _\mu }{\theta _\nu }}} &= Tr\left[ {{\partial _{{\theta _\mu }}}\hat \rho {\cal L}{{_{{\theta _\nu }}^R}^\dag }} \right]  \\ \notag&
	\mathop { =\joinrel=}\limits^{(\ref{27})}{{\cal L}^{R\left( 0 \right)*}}Tr\left[ {{\partial _{{\theta _\mu }}}\hat \rho } \right] + {\cal L}_l^{R\left( 1 \right)*}Tr\left[ {{\partial _{{\theta _\mu }}}\hat \rho  \hspace{0.1cm} {{\hat r}_l}} \right] + {\cal L}{_{kj}^{R\left( 2 \right)*}}Tr\left[ {{\partial _{{\theta _\mu }}}\hat \rho \hspace{0.1cm} {{\hat r}_k}{{\hat r}_j}} \right] \\ \notag&
	\mathop {=\joinrel=} \limits^{(\ref{B2})}{{\cal L}^{R\left( 0 \right)*}}{\left. {{\partial _{{\theta _\mu }}}Tr\left[ {\hat D \hat \rho } \right]} \right|_{\tilde r = 0}} + {\left. {{\cal L}_l^{R\left( 1 \right)*}{\partial _{{\theta _\mu }}} Tr\left[ {\hat D \hspace{0.1cm} \hat \rho \hspace{0.1cm} {{\hat r}_l}} \right]} \right|_{\tilde r = 0}} + {\left. {{\cal L}{{_{kj}^{R\left( 2 \right)}}^*}{\partial _{{\theta _\mu }}}Tr\left[ {\hat D \hspace{0.1cm} \hat \rho \hspace{0.1cm} {{\hat r}_k} \hspace{0.1cm} {{\hat r}_j}} \right]} \right|_{\tilde r=0}}\\ \notag&
	\mathop { =\joinrel=}\limits^{(\ref{A2})} {{\cal L}^{R\left( 0 \right)*}}{\left. {{\partial _{{\theta _\mu }}}{\chi _{\hat \rho }}} \right|_{\tilde r = 0}} + {\left. {{\cal L}_l^{R\left( 1 \right)*}\left( { - i{\partial _{{{\tilde r}_l}}} - \frac{1}{2}{\Omega _{ll'}}{{\tilde r}_{l'}}} \right){\partial _{{\theta _\mu }}}{\chi _{\hat \rho }}} \right|_{\tilde r = 0}} +  {\left. {{\cal L}{{_{kj}^{R\left( 2 \right)}}^*}\left( { - i{\partial _{{{\tilde r}_j}}} - \frac{1}{2}{\Omega _{jj'}}{{\tilde r}_{j'}}} \right)\left( { - i{\partial _{{{\tilde r}_k}}} - \frac{1}{2}{\Omega _{kk'}}{{\tilde{r}}_{k'}}} \right){\partial _{{\theta _\mu }}}{\chi _{\hat \rho }}} \right|_{\tilde r = 0}}.
	\end{align}
	Replacing ${\partial _{{\theta _\mu }}}{\chi _{\hat \rho }}$ by the corresponding expression in (\ref{A4}), one has
	\begin{align*}
		{{\cal F}_{{\theta _\mu }{\theta _\nu }}}&={\mathcal{L}^{R\left( 0 \right)*}}{\left. {\left( {i{\mkern 1mu} {{\tilde r}_v}{\partial _{{\theta _\mu }}}{d_v} - \frac{1}{4}{\partial _{{\theta _\mu }}}{\sigma _{pm}}{{\tilde r}_p}{{\tilde r}_m}} \right){\chi _{\hat \rho }}} \right|_{\tilde r = 0}} + {\left. {\mathcal{L}_l^{R\left( 1 \right)*}\left( { - i{\partial _{{{\bar r}_l}}} - \frac{1}{2}{\Omega _{ll'}}{{\tilde r}_{l'}}} \right)\left( {i{\mkern 1mu} {{\tilde r}_v}{\partial _{{\theta _\mu }}}{d_v} - \frac{1}{4}{\partial _{{\theta _\mu }}}{\sigma _{pm}}{{\tilde r}_p}{{\tilde r}_m}} \right){\chi _{\hat \rho }}} \right|_{\tilde r = 0}} + \\ \notag& \hspace*{4cm}
	{\rm{  \mathcal{L}}}_{kj}^{R\left( 2 \right)*}{\left. {\left( { - i{\partial _{{{\tilde r}_j}}} - \frac{1}{2}{\Omega _{jj'}}{{\tilde r}_{j'}}} \right)\left( { - i{\partial _{{{\tilde r}_k}}} - \frac{1}{2}{\Omega _{kk'}}{{\tilde r}_{k'}}} \right)\left( {i{\mkern 1mu} {{\tilde r}_v}{\partial _{{\theta _\mu }}}{d_v} - \frac{1}{4}{\partial _{{\theta _\mu }}}{\sigma _{pm}}{{\tilde r}_p}{{\tilde r}_m}} \right){\chi _{\hat \rho }}} \right|_{\tilde r = 0}}.
	\end{align*}
	 Using the expression of ${\partial _{{{\tilde r}_l}}}{\chi _{\hat \rho }}$ given by (\ref{A4}), one gets
	\begin{equation}
	{\cal F}_{{\theta _\mu }{\theta _\nu }} = {\cal L}_l^{R\left( 1 \right)*}{\partial _{{\theta _\mu }}}{d_l} + {\cal L}_{kj}^{R\left( 2 \right)*}\left( {\frac{1}{2}{\partial _{{\theta _\mu }}}{\sigma _{kj}} + 2\,{\partial _{{\theta _\mu }}}{d_k}\,{d_j}} \right).
	\end{equation}
	Using the matrix representation, one finds
	\begin{equation}
	{{\cal F}_{{\theta _\mu }{\theta _\nu }}} = \frac{1}{2}Tr\left[ {{\partial _{{\theta _\mu }}}\sigma \mathcal{L}_{{\theta _\nu }}^{R\left( 2 \right)\dag }} \right] + \mathcal{L}_{{\theta _\nu }}^{R\left( 1 \right)\dag }{\partial _{{\theta _\mu }}}{\bf{d}} + 2{\partial _{{\theta _\mu }}}{{\bf{d}}^T}\mathcal{L}_{{\theta _\nu }}^{R\left( 2 \right)\dag }{\bf{d}}.
	\end{equation}
	Using Eq. (\ref{40}) and replacing $\mathcal{L}_{{\theta _\nu }}^{R\left( 2 \right)}$ by its expression given in (\ref{A11}), we find the expression of RLD quantum Fisher information matrix  
	\begin{equation}
	{\mathcal{F}_{{\theta _\mu }{\theta _\nu }}} = \frac{1}{2}\mathtt{vec}{\left[ {{\partial _{{\theta _\mu }}}\sigma } \right]^\dag }{\left( {\Gamma  ^\dag \otimes {\Gamma }} \right)^{+}}\mathtt{vec}\left[ {{\partial _{{\theta _\nu }}}\sigma } \right] + 2{\partial _{{\theta _\mu }}}{\mathbf{d}^T}\Gamma  ^{+}{\partial _{{\theta _\nu }}}\mathbf{d}.
	\end{equation}
	
	\section{Symmetric Logarithmic Derivative (SLD) \label{App:C}}
Analogously, to find the explicit expression of the SLD quantum Fisher information matrix, it is necessary to determine first the expression of the corresponding symmetric logarithmic derivative (SLD) operator. In this case, we consider the quadratic form of SLD given by (\ref{33}) and one has to determiner the expression of ${L^S}^{\left( 0 \right)}$,  $L^{S\left( 1 \right)}$ and $L^{S\left( 2 \right)}$. To do this, one starts with
	\begin{align}
	{\partial _{{\theta _\mu }}}{\chi _{\hat \rho }} &= Tr\left[ {\hat D \hspace{0.1cm}{\partial _{{\theta _\mu }}}\hat \rho } \right]\\ \notag&
	\mathop  { =\joinrel=} \limits^{\left( {\ref{14}} \right)} \frac{1}{2}\left( {Tr\left[ {\hat D \hat \rho \hat L_{{\theta _\mu }}^S} \right] + Tr\left[ {\hat D \hat L_{{\theta _\mu }}^S\hat \rho } \right]} \right)\\ \notag&
	\mathop { =\joinrel=} \limits^{\left( {\ref{33}} \right)} \frac{1}{2}\left( {Tr\left[ {\hat D \hat \rho \left( {{L^S}^{\left( 0 \right)} + L{{_l^S}^{\left( 1 \right)}}{{\hat r}_l} + L{{_{jk}^S}^{\left( 2 \right)}}{{\hat r}_j}{{\hat r}_k}} \right)} \right] + Tr\left[ {\hat D \left( {{L^S}^{\left( 0 \right)} + L{{_l^S}^{\left( 1 \right)}}{{\hat r}_l} + L{{_{jk}^S}^{\left( 2 \right)}}{{\hat r}_j}{{\hat r}_k}} \right)\hat \rho } \right]} \right)\\ \notag&
	 {=\joinrel=} {L^S}^{\left( 0 \right)}Tr\left[ {\hat D \hat \rho } \right] + \frac{1}{2}L_l^{S\left( 1 \right)}Tr\left[ {\hat D \left( {\hat \rho \hspace{0.1cm} {{\hat r}_l} + {{\hat r}_l} \hspace{0.1cm} \hat \rho } \right)} \right] + \frac{1}{2}L_{jk}^{S\left( 2 \right)}Tr\left[ {\hat D \left( {{{\hat r}_j}{{\hat r}_k}\hat \rho  + \hspace{0.1cm}\hat \rho \hspace{0.1cm} {{\hat r}_j}{{\hat r}_k}} \right)} \right]\\ \notag&
	\mathop  { =\joinrel=} \limits^{(\ref{A3})} {L^S}^{\left( 0 \right)}{\chi _{\hat \rho }} + L_l^{S\left( 1 \right)}\left( { - i{\mkern 1mu} {\mkern 1mu} {\partial _{{{\tilde r}_l}}}{\chi _{\hat \rho }}} \right) + \frac{1}{2}L_{jk}^{S\left( 2 \right)}\left( { - 4{\mkern 1mu} {\partial _{{{\tilde r}_j}}}{\mkern 1mu} {\mkern 1mu} {\partial _{{{\tilde r}_k}}} + {\Omega _{jj'}}{{\tilde r}_{j'}}{\Omega _{kk'}}{{\tilde r}_{k'}}} \right){\chi _{\hat \rho }}.
	\end{align}
	Replacing the results of (\ref{A4}) in the last equation, one finds
	\begin{align}\label{C2}
	\left( {2i\,{{\tilde r}_v}{\partial _{{\theta _\mu }}}{d_v} - \frac{1}{2}{\partial _{{\theta _\mu }}}{\sigma _{lm}}{{\tilde r}_l}{{\tilde r}_m}} \right){\chi _{\hat \rho }} = & 2{L^S}^{\left( 0 \right)}{\chi _{\hat \rho }} + L_l^{S\left( 1 \right)}\left( {2\,{d_l} + i\hspace{0.1cm}{\sigma _{ll'}}{{\tilde r}_{l'}}} \right){\chi _{\hat \rho }} + \frac{1}{2}L_{jk}^{S\left( 2 \right)}{\Omega _{jj'}}{{\tilde r}_{j'}}{\Omega _{kk'}}{{\tilde r}_{k'}}\,{\chi _{\hat \rho }} -\\ \notag& \hspace{1cm}
	 L_{jk}^{S\left( 2 \right)}\left( {2\left( {i{d_j} - \frac{1}{2}{\sigma _{jj'}}{{\tilde r}_{j'}}} \right)\left( {i{d_k} - \frac{1}{2}{\sigma _{kk'}}{{\tilde r}_{k'}}} \right) - {\sigma _{jk}}} \right){\chi _{\hat \rho }}. 
	\end{align}
	The identification of second-order terms of Eq. (\ref{C2}) leads to:
\begin{equation}
	{\partial _{{\theta _\mu }}}{\sigma _{lm}} = L_{jk}^{S\left( 2 \right)}{\sigma _{jj'}}{\sigma _{kk'}} - L_{jk}^{S\left( 2 \right)}{\Omega _{jj'}}{\Omega _{kk'}},
\end{equation}
and the matrix representation form is
	\begin{equation}
{\partial _{{\theta _\mu }}}\sigma  = \sigma L_{{\theta _\mu }}^{S\left( 2 \right)}\sigma  - \Omega L_{{\theta _\mu }}^{S\left( 2 \right)}\Omega.
\end{equation}
Using the Eq. (\ref{A10}), one obtains
\begin{equation}
\mathtt{vec}\left[ {L_{{\theta _\mu }}^{S\left( 2 \right)}} \right] = {\left( {{\sigma ^\dag} \otimes \sigma  + \Omega  \otimes \Omega } \right)^{+}}\mathtt{vec}\left[ {{\partial _{{\theta _\mu }}}\sigma } \right]. \label{C6}
\end{equation}
The identification of first-order terms  of Eq. (\ref{C2}) leads to :
	\begin{equation}
	2{\partial _{{\theta _\mu }}}{d_v} = L_l^{S\left( 1 \right)}{\sigma _{ll'}} + L_{jk}^{S\left( 2 \right)}\left( {{d_j}{\sigma _{kk'}} + {d_k}{\sigma _{jj'}}} \right),
	\end{equation}
and the corresponding matrix form is
	\begin{equation}
	L_{{\theta _\mu }}^{S\left( 1 \right)} = 2{\sigma ^{-1}}{\partial _{{\theta _\mu }}}{\bf{d}} - 2L_{{\theta _\mu }}^{S\left( 2 \right)}{\bf{d}}.
	\end{equation}
	The identification of zero-order terms  of Eq. (\ref{C2}) leads to :
	\begin{equation}
	2{L^S}^{\left( 0 \right)} + 2L_l^{S\left( 1 \right)}{d_l} + L_{jk}^{S\left( 2 \right)}{\sigma _{jk}} + 2L_{jk}^{S\left( 2 \right)}{d_j}{d_k} = 0,
	\end{equation}
which takes the following matrix form
	\begin{equation}
	L_{{\theta _\mu }}^{S\left( 0 \right)} =  - \frac{1}{2}Tr\left[ {L_{{\theta _\mu }}^{S\left( 2 \right)}\sigma } \right] - L_{{\theta _\mu }}^{S\left( 1 \right)T}{\bf{d}} - {{\bf{d}}^T}L_{{\theta _\mu }}^{S\left( 2 \right)}{\bf{d}}.
	\end{equation}
	\section{SLD Quantum Fisher Information Matrix (QFIM) \label{App:D}}
	 To find the formula of the SLD quantum Fisher information matrix, we insert the expression of the symmetric logarithmic derivative operator (SLD) which is obtained in Appendix (\ref{App:C}) into Eq. (\ref{16}). This equation can be also written as
	\begin{equation}
	{H_{{\theta _\mu }{\theta _\nu }}}= \frac{1}{2}Tr\left[ {{\partial _{{\theta _\mu }}}\hat \rho \,\hat L_{{\theta _\nu }}^S} \right].
	\end{equation}
	Using the property of characteristic function given by (\ref{B2}), one gets
	\begin{align}
	{H_{{\theta _\mu }{\theta _\nu }}} &= \frac{1}{2}Tr\left[ {{\partial _{{\theta _\mu }}}\hat \rho \,\hat L_{{\theta _\nu }}^S} \right]
	\\ \notag&
	\mathop { =\joinrel= }\limits^{\left( {\ref{33}} \right)} \frac{1}{2}Tr\left[ {{\partial _{{\theta _\mu }}}\hat \rho \left( {{L^S}^{\left( 0 \right)} + L_l^{S\left( 1 \right)}{{\hat r}_l} + L_{jk}^{S\left( 2 \right)}{{\hat r}_j}{{\hat r}_k}} \right)} \right] \\ \notag&
	{=\joinrel=} \frac{1}{2}{L^S}^{\left( 0 \right)}Tr\left[ {{\partial _{{\theta _\mu }}}\hat \rho } \right] + L_l^{S\left( 1 \right)}Tr\left[ {{\partial _{{\theta _\mu }}}\hat \rho \hspace{0.1cm} {{\hat r}_l}} \right] + L_{jk}^{S\left( 2 \right)}Tr\left[ {{\partial _{{\theta _\mu }}}\hat \rho \hspace{0.1cm}{{\hat r}_j}\hspace{0.1cm}{{\hat r}_k}} \right] \\ \notag&
	\mathop{=\joinrel=}\limits^{\left({\ref{B2}}\right)}\frac{1}{2}{L^S}^{\left( 0 \right)}{\left. {{\partial _{{\theta _\mu }}}{\chi _{\hat \rho }}} \right|_{\tilde r = 0}} + L_l^{S\left( 1 \right)}{\left. {Tr\left[ {\hat D \hspace{0.1cm}\hat \rho \hspace{0.1cm} {{\hat r}_l}} \right]{\partial _{{\theta _\mu }}}{\chi _{\hat \rho }}} \right|_{\tilde r= 0}} + {\left. {L_{jk}^{S\left( 2 \right)}Tr\left[ {\hat D \hspace{0.1cm}\hat \rho \hspace{0.1cm} {{\hat r}_j}\hspace{0.1cm}{{\hat r}_k}} \right]{\partial _{{\theta _\mu }}}{\chi _{\hat \rho }}} \right|_{\tilde r = 0}} \\ \notag&
	\mathop{=\joinrel=}\limits^{\left( {\ref{A2}} \right)} \frac{1}{2}{L^S}^{\left( 0 \right)}{\left. {{\partial _{{\theta _\mu }}}{\chi _{\hat \rho }}} \right|_{\tilde r = 0}} + L_l^{S\left( 1 \right)}{\left. {\left( { - i{\partial _{{{\tilde r}_l}}} - \frac{1}{2}{\Omega _{ll'}}{{\tilde r}_{l'}}} \right){\partial _{{\theta _\mu }}}{\chi _{\hat \rho }}} \right|_{\tilde r = 0}} + {\left. {L_{jk}^{S\left( 2 \right)}\left( { - i{\partial _{{{\tilde r}_j}}} - \frac{1}{2}{\Omega _{jj'}}{{\tilde r}_{j'}}} \right)\left( { - i{\partial _{{{\tilde r}_k}}} - \frac{1}{2}{\Omega _{kk'}}{{\tilde r}_{k'}}} \right){\partial _{{\theta _\mu }}}{\chi _{\hat \rho }}} \right|_{\tilde r = 0}}\\ \notag&
	\mathop  {=\joinrel=} \limits^{\left( {\ref{A4}} \right)} \frac{1}{2}{L^S}^{\left( 0 \right)}{\left. {\left( {i\,{{\tilde r}_v}{\partial _{{\theta _\mu }}}{d_v} - \frac{1}{4}{\partial _{{\theta _\mu }}}{\sigma _{pm}}{{\tilde r}_p}{{\tilde r}_m}} \right){\chi _{\hat \rho }}} \right|_{\bar r = 0}} + L_l^{S\left( 1 \right)}{\left. {\left( { - i{\partial _{{{\tilde r}_l}}} - \frac{1}{2}{\Omega _{ll'}}{{\tilde r}_{l'}}} \right)\left( {i\,{{\tilde r}_v}{\partial _{{\theta _\mu }}}{d_v} - \frac{1}{4}{\partial _{{\theta _\mu }}}{\sigma _{pm}}{{\tilde r}_p}{{\tilde r}_m}} \right){\chi _{\hat \rho }}} \right|_{\tilde r = 0}} +\\ \notag& \hspace{4cm}{\left. {L_{jk}^{S\left( 2 \right)}\left( { - i{\partial _{{{\tilde r}_j}}} - \frac{1}{2}{\Omega _{jj'}}{{\tilde r}_{j'}}} \right)\left( { - i{\partial _{{{\tilde r}_k}}} - \frac{1}{2}{\Omega _{kk'}}{{\tilde r}_{k'}}} \right)\left( {i\,{{\tilde r}_v}{\partial _{{\theta _\mu }}}{d_v} - \frac{1}{4}{\partial _{{\theta _\mu }}}{\sigma _{pm}}{{\tilde r}_p}{{\tilde r}_m}} \right){\chi _{\hat \rho }}} \right|_{\tilde r = 0}}.
	\end{align}
Evaluating  the last equation when $\mathbf{\tilde{r}}=0$, one gets
	\begin{equation}
	{H_{{\theta _\mu }{\theta _\nu }}}= L_l^{S\left( 1 \right)}{\partial _{{\theta _\mu }}}{d_l} + \frac{1}{2}L_{jk}^{S\left( 2 \right)}{\partial _{{\theta _\mu }}}{\sigma _{jk}} + 2L_{jk}^{S\left( 2 \right)}{\partial _{{\theta _\mu }}}{d_j}{d_k},
	\end{equation}
which rewrites  in the matrix representation as
	\begin{equation}
	{H_{{\theta _\mu }{\theta _\nu }}} = {\partial _{{\theta _\mu }}}{\mathbf{d}^T}L_{{\theta _\nu }}^{S\left( 1 \right)} + \frac{1}{2}Tr\left[ {{\partial _{{\theta _\mu }}}\sigma L_{{\theta _\nu }}^{S\left( 2 \right)}} \right] + 2{\partial _{{\theta _\mu }}}{\mathbf{d}^T}L_{{\theta _\nu }}^{S\left( 2 \right)}\mathbf{d}.
	\end{equation}
	Using Eq. (\ref{40}), and replacing $L_{{\theta _\nu }}^{S\left( 2 \right)}$ by its expression given by (\ref{C6}), one obtains the following SLD quantum Fisher information matrix 
	\begin{equation}
	{H_{{\theta _\mu }{\theta _\nu }}} = \frac{1}{2}\mathtt{vec}{\left[ {{\partial _{{\theta _\mu }}}\sigma } \right]^\dag }{{\mathcal M}^{+}}\mathtt{vec}\left[ {{\partial _{{\theta _\nu }}}\sigma } \right] + 2{\partial _{{\theta _\mu }}}{{\bf{d}}^T} \hspace{0.1cm} \sigma ^{-1} \hspace{0.1cm}{\partial _{{\theta _\nu }}}{\bf{d}},
	\end{equation}
	where $\mathcal{M} = \left( {{\sigma ^\dag} \otimes \sigma  + \Omega  \otimes \Omega } \right)$.
\end{widetext}

\end{document}